\documentclass[onecolumn, 10pt]{IEEEtran}
\usepackage{amsmath}
\usepackage{amssymb}
\usepackage{amsfonts}
\usepackage{cite}
\usepackage{ifthen}
\usepackage{epsfig}
\usepackage{array}
\usepackage{enumerate}
\usepackage{subfig}
\usepackage{algorithm}
\usepackage[all]{xy}
\usepackage{algpseudocode}
\newcommand{\subparagraph}{}

\usepackage{times}
\usepackage{tikz}
\usepackage{paralist}
\usepackage{float}
\usetikzlibrary{shapes,positioning,arrows,automata}
\usepackage{caption}
\newtheorem{definition}{Definition} 
\newtheorem{thm}{Theorem}
\newtheorem{theorem}{Theorem}

\newtheorem{lem}[thm]{Lemma}

\newtheorem{remark}{Remark}
\newtheorem{example}{Example}

\newcommand{\beq}{\begin{equation}}
\newcommand{\eeq}{\end{equation}}
\newcommand{\bea}{\begin{eqnarray}}
\newcommand{\eea}{\end{eqnarray}}
\newcommand{\bean}{\begin{eqnarray*}}
\newcommand{\eean}{\end{eqnarray*}}
\newcommand{\bit}{\begin{itemize}}
\newcommand{\eit}{\end{itemize}}
\newcommand{\ben}{\begin{enumerate}}
\newcommand{\een}{\end{enumerate}}
\newcommand{\blem}{\begin{lem}}
\newcommand{\elem}{\end{lem}}
\newcommand{\bthm}{\begin{thm}}
\newcommand{\ethm}{\end{thm}}
\newcommand{\bpf}{\begin{IEEEproof}}
\newcommand{\epf}{\end{IEEEproof}}
\newcommand{\plotwidth}{.45\textwidth}
\newcommand{\demoWidth}{.13\textwidth}
\newcommand{\demoGap}{.05\textwidth}
\newcommand{\modulo}{\text{\,\textrm{mod}\,}}
\newcommand{\comment}[1]{}
\newcommand{\supth}{^{\textrm{th}}}

\newcommand{\upseck}{\textsuperscript{\small \lowercase{k}}}
\newcommand{\MkMn}{M\upseck/M/\lowercase{n}}
\newcommand{\BigOmega}[1]{\ensuremath{\operatorname{\Omega}\bigl(#1\bigr)}}

\newdir{halfFive}{!/5pt/@{ }*:(0,0)@^{ }*:(0,0)@_{ }}
\newdir{halfTen}{!/10pt/@{ }*:(0,0)@^{ }*:(0,0)@_{ }}
\newdir{halfTwenty}{!/20pt/@{ }*:(0,0)@^{ }*:(0,0)@_{ }}
\newdir{halfThirty}{!/30pt/@{ }*:(0,0)@^{ }*:(0,0)@_{ }}
\newdir{halfForty}{!/40pt/@{ }*:(0,0)@^{ }*:(0,0)@_{ }}
\newdir{halfFifty}{!/50pt/@{ }*:(0,0)@^{ }*:(0,0)@_{ }}
\newdir{halfSixty}{!/60pt/@{ }*:(0,0)@^{ }*:(0,0)@_{ }}
\newdir{halfSeventy}{!/70pt/@{ }*:(0,0)@^{ }*:(0,0)@_{ }}
\newdir{halfHundred}{!/100pt/@{ }*:(0,0)@^{ }*:(0,0)@_{ }}

\newcommand{\algostate}{\State\hspace{-.3cm}}
\newcommand{\algoindent}{\hspace{.3cm}}

\emergencystretch=\maxdimen 

\title{The MDS Queue: Analysing the Latency Performance of Erasure Codes}
\author{Nihar B. Shah, Kangwook Lee, Kannan Ramchandran\\Dept. of Electrical Engineering and Computer Sciences\\ University of California, Berkeley\\\{nihar,\,kw1jjang,\,kannanr\}@eecs.berkeley.edu\thanks{N. B. Shah was supported by a Berkeley Fellowship and K. Lee by a KFAS Fellowship. This work was also supported in part by NSF grant CCF-1116404.}}

\begin{document}
\maketitle
\thispagestyle{empty}

\begin{abstract}
In order to scale economically, data centers are increasingly evolving their data storage methods from the use of simple data replication to the use of more powerful erasure codes, which provide the same level of reliability as replication but at a significantly lower storage cost. In particular, it is well known that Maximum-Distance-Separable (MDS) codes, such as Reed-Solomon codes, provide the maximum storage efficiency. While the use of codes for providing improved reliability in archival storage systems, where the data is less frequently accessed (or so-called ``cold data''), is well understood, the role of codes in the storage of more frequently accessed and active ``hot data'', where latency is the key metric, is less clear. 

In this paper, we study data storage systems based on MDS codes through the lens of queueing theory, and term this the ``MDS queue.'' We analytically characterize the (average) latency performance of MDS queues, for which we present insightful scheduling policies that form upper and lower bounds to performance, and are observed to be quite tight. Extensive simulations are also provided and used to validate our theoretical analysis. We also employ the framework of the MDS queue to analyse different methods of performing so-called degraded reads (reading of partial data) in distributed data storage.
\end{abstract}

\section{Introduction}\label{sec:intro}
Two of the primary objectives of a storage system are to provide reliability and availability of the stored data: the system must ensure that data is not lost even in the presence of individual component failures, and must be easily and quickly accessible to the user whenever required. The classical means of providing reliability is to employ the strategy of replication, wherein identical copies of the (entire) data are stored on multiple servers. However, this scheme is not very efficient in terms of the storage space utilization. The exponential growth in the amount of data being stored today makes storage an increasingly valuable resource, and has motivated data-centers today to increasingly turn to the use of more efficient erasure codes~\cite{hdfs_codes_blog,cleversafe_erasure_codes,ford2010availability,huang2012erasure,rashmi2013hotstorage}.

The most popular, and also most efficient storage codes are the Maximum-Distance-Separable (MDS) codes, e.g., Reed-Solomon codes. An MDS code is typically associated to two parameters $n$ and $k$. Under an $(n,k)$ MDS code, a file is encoded and stored in $n$ servers such that (a) the data stored in \textit{any} $k$ of these $n$ servers suffice to recover the entire file, and (b) the storage space required at each server is $\frac{1}{k}$ of the size of the original file.~\footnote{A more generic definition of an MDS code is that it is a code that satisfies the `Singleton bound'~\cite{SloaneBook}.}

\begin{figure}[t!]
\centering
\includegraphics[width=\plotwidth]{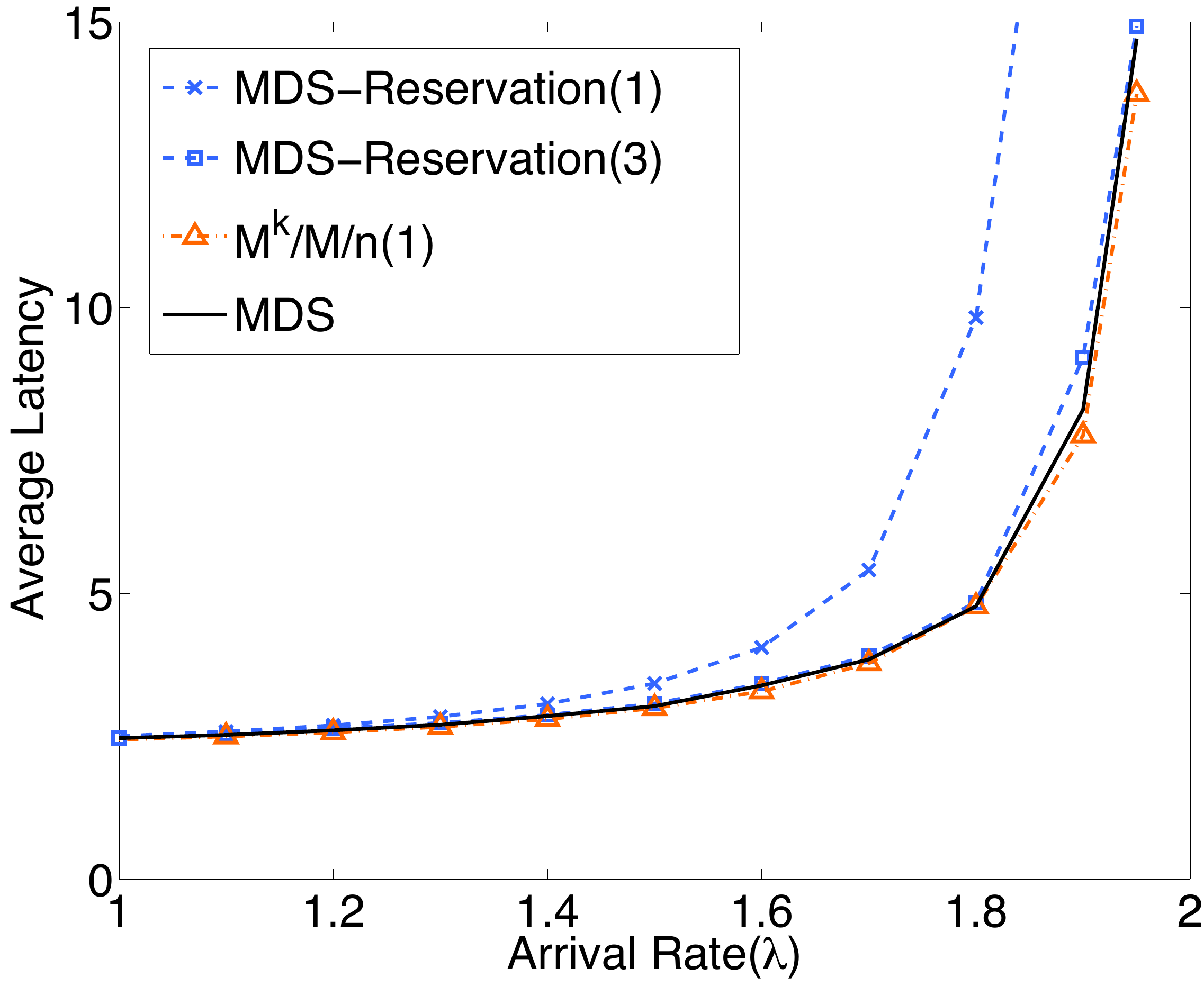}
\caption{The average latency of a system using an MDS code with $n=10$ and $k=5$. The service of each job is assumed to be drawn from an exponential distribution with rate $\mu=1$. The curve titled `MDS' corresponds to simulations of the exact coded system. Also plotted are the analytically computed latencies of the lower bounds (MDS-Reservation(t) queues) and upper bounds (\MkMn(t) queues) presented in this paper. We also confirmed that these analytically computed latencies closely match the simulated performances of the corresponding scheduling policies.}
\label{fig:preview}
\end{figure}

While the reliability properties of erasure codes are very well understood, much less is known about their latency performance. In this paper, we study coded data storage systems based on MDS codes through the lens of queueing theory. We term the queue resulting from the use of codes that allow for recovery of the data from any $k$ of the $n$ nodes as ``the {MDS queue}''. To understand this queueing-theoretic perspective, consider a simple example with $n=4$ and $k=2$. Several files $\{F_i\}$ are to be stored in the $4$ servers in a manner that no data is lost upon  failure of any $(n-k)=2$ of the $n=4$ servers. This is achieved via a $(4,\ 2)$ MDS code under which, each file is partitioned into two halves $F_i = [f_{i,1}~f_{i,2}]$, and the $4$ servers store 
$f_{i,1}$, $f_{i,2}$, $(f_{i,1} + f_{i,2})$, and $(f_{i,1} + 2f_{i,2})$ respectively for all $i$. 
Requests for reading individual files arrive as a stochastic process, which are buffered and served by the system, and the resulting queue is termed an MDS queue (this shall be formalized later in the paper).

An exact analysis of the MDS queue is hard in general: a Markov chain representation of the MDS queue has a state space that is infinite in at-least $k$ dimensions, and furthermore, the transitions are tightly coupled across the $k$ dimensions. In this paper, we present insightful scheduling policies that provide upper and lower bounds to the performance of the MDS queue, and these are observed to be quite tight. Using these bounds we analytically characterize the (average) latency performance of MDS queues, as illustrated in Fig.~\ref{fig:preview}. 

The lower bounds (the `{MDS-Reservation(t)}' scheduling policies) and the upper bounds (the `{\MkMn(t)}' scheduling policies) presented in this paper are both indexed by a parameter `t'. An increase in the value of t results in tighter bounds, but also increases the complexity of analysis. Furthermore, both classes of scheduling policies converge to the MDS scheduling policy as t$ \rightarrow \infty$. However, we observe that the performance of the MDS-Reservation(t) queue is very close to that of the MDS queue for very small values of t (as small as $\textrm{t}=3$), and the performance of the upper bounds \MkMn(t) closely follow that of the MDS queue for values of t as small as $\textrm{t}=1$. This can be observed in Fig.~\ref{fig:preview}. The MDS-Reservation(t) scheduling policies presented here are themselves practical alternatives to the MDS scheduling policy, since they require maintenance of a smaller state, while offering highly comparable performance.

We also consider the problem of degraded reads (i.e., reading of partial data) in distributed storage systems, that has recently attracted considerable interest in the coding-theory community. We employ the framework of the MDS queue to understand and compare, from a queueing theoretic viewpoint, different methods of performing degraded reads.

The rest of the paper is organized as follows. Section~\ref{sec:definition} presents the MDS queue system model. Section~\ref{sec:literature} discusses related literature. Section~\ref{sec:approach} describes the general approach and the notation followed in the paper. Section~\ref{sec:lower} presents the MDS-Reservation(t) queues that lower bound the performance of the MDS queue. Section~\ref{sec:upper} presents the \MkMn(t) queues that upper bound the performance.  Section~\ref{sec:analysis} presents analyses and comparisons of these queues. Section~\ref{sec:future} presents conclusions and discusses open problems. The appendix contains proofs of the theorems presented in the paper.

\section{The MDS Queue System Model}\label{sec:definition}
We shall now describe a queueing theoretic model of a system employing an MDS code. 
As discussed previously, under an MDS code, a file can be retrieved by downloading data from any $k$ of the servers. We model this by treating each request for reading a file as a \textit{batch} of $k$ \textit{jobs}. The $k$ jobs of a batch represent reading of $k$ encodings of the file from $k$ servers. A batch is considered as served when $k$ of its jobs have been served. For instance, in the example of the $(n=4,\ k=2)$ system of Section~\ref{sec:intro}, a request for reading file $F_i$ (for some $i$) is treated as a batch of two jobs. To serve this request, the two jobs may be served by any two of the four servers; for example, if the two jobs are served by servers $2$ and $3$, then they correspond to reading $f_{i,2}$ and $(f_{i,1}+f_{i,2})$ respectively, which suffice to obtain $F_i$. We assume homogeneity among files and among servers: the files are of identical size, and the $n$ servers have identical performance characteristics.

\begin{figure*}[bt]
\centering
\subfloat[]{
\includegraphics[width=\demoWidth]{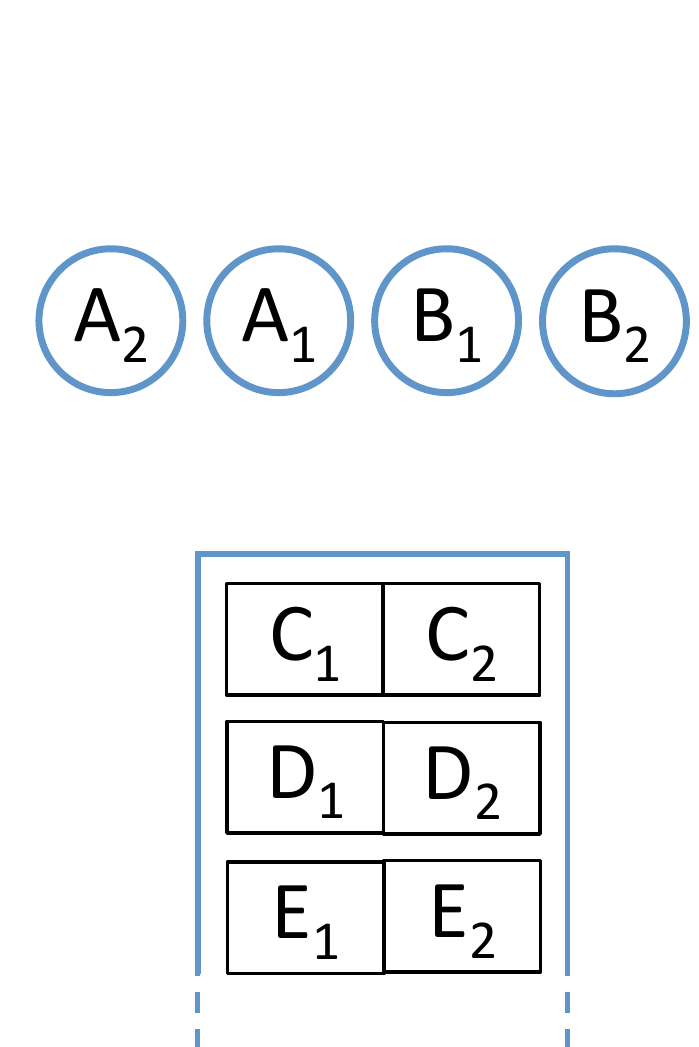}
\label{fig:MDSqueue_working_a}
}
\hspace{\demoGap}
\subfloat[]{
\includegraphics[width=\demoWidth]{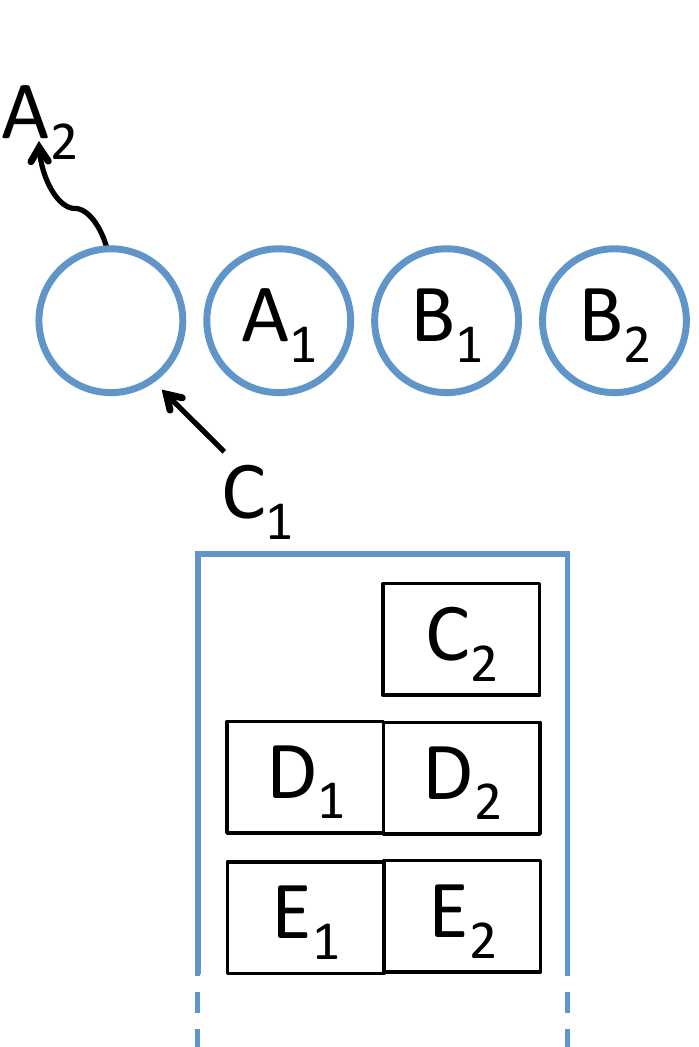}
\label{fig:MDSqueue_working_b}
}
\hspace{\demoGap}\subfloat[]{
\includegraphics[width=\demoWidth]{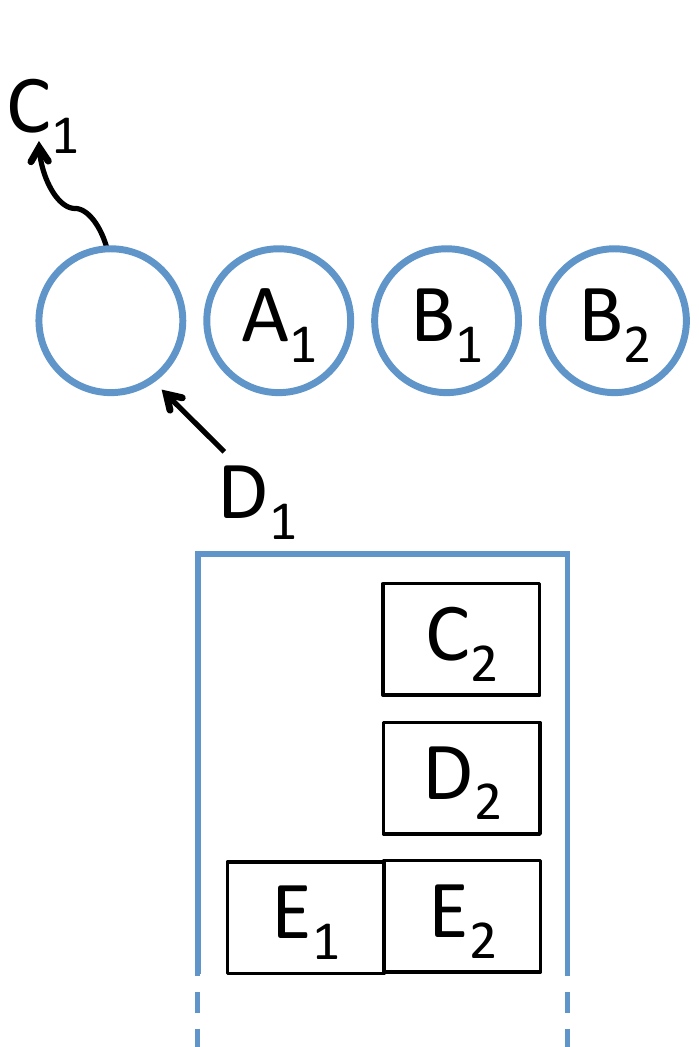}
\label{fig:MDSqueue_working_c}
}
\hspace{\demoGap}\subfloat[]{
\includegraphics[width=\demoWidth]{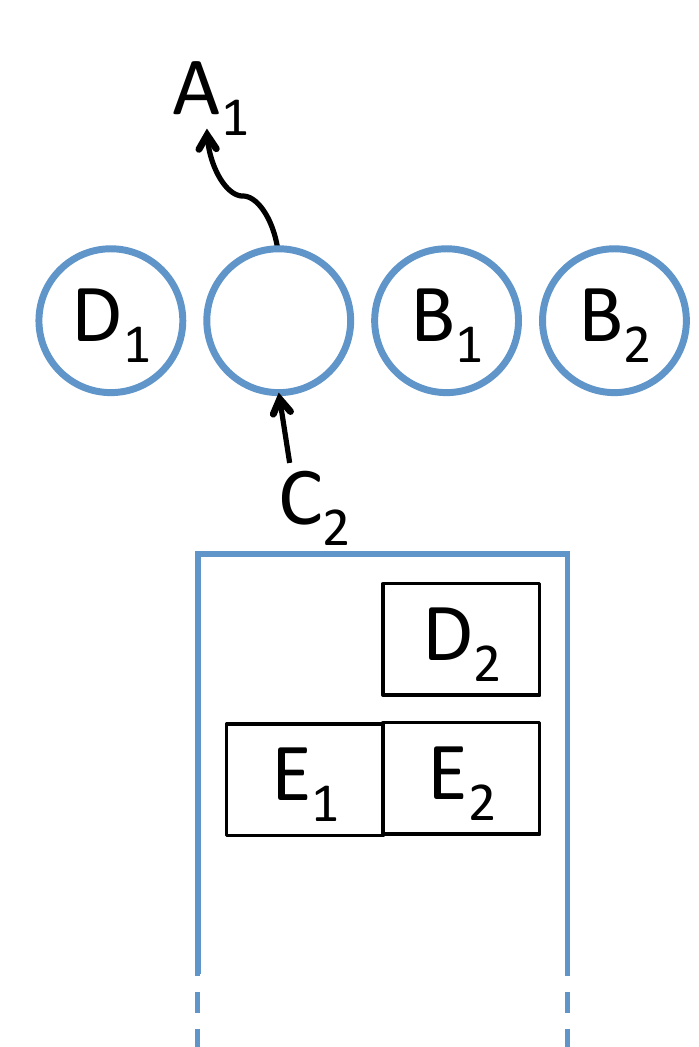}
\label{fig:MDSqueue_working_d}
}
\caption{Functioning of the MDS queue.}
\label{fig:MDSqueue_working}
\end{figure*}

\begin{definition}[MDS queue]
An MDS queue is associated to four parameters $(n,k)$ and $[\lambda,\mu]$.\begin{itemize}
\itemsep0em
\item There are $n$ identical servers
\item Requests enter into a (common) buffer of infinite capacity
\item Requests arrive as a Poisson process with rate $\lambda$
\item Each request comprises a \textbf{batch} of $k$ \textbf{jobs}
\item Each of the $k$ jobs in a batch can be served by an \textbf{arbitrary} set of $k$ \textbf{distinct} servers
\item The service time for a job at any server is exponentially distributed with rate $\mu$, independent of all else
\item The jobs are processed in order, i.e., among all the waiting jobs that an idle server is allowed to serve, it serves the one which had arrived the earliest.
\end{itemize}
\end{definition}
Algo.~\ref{alg:MDS} formalizes the scheduling policy of the MDS queue.
\begin{algorithm}
\begin{algorithmic}
\algostate \textbf{On} arrival of a batch
\algostate \algoindent Assign as many of its jobs as possible to idle servers
\algostate \algoindent Append remaining jobs (if any) as new batch at end of buffer 
\algostate \textbf{On} departure from a server (say, server $s$)
\algostate \algoindent  \textbf{If} {$\exists$ at least one batch in the buffer such that no job of this batch has been served by $s$}
\algostate \algoindent \algoindent Among all such batches, find batch that arrived earliest
\algostate \algoindent \algoindent Assign a job from this batch to $s$
\end{algorithmic}
\caption{MDS scheduling policy}
\label{alg:MDS}
\end{algorithm}

The following example illustrates the functioning of the MDS scheduling policy and the resultant MDS queue.
\begin{example}
Consider the MDS$(n=4,\ k=2)$ queue, as depicted in Fig.~\ref{fig:MDSqueue_working}. Here, each request comes as a batch of $k=2$ jobs, and hence we denote each batch (e.g., $A$, $B$, etc.) as a pair of jobs ($\{A_1,A_2\}$, $\{B_1,B_2\}$, etc.). The two jobs in a batch need to be served by (any) two distinct servers. Denote the four servers (from left to right) as servers $1$, $2$, $3$ and $4$.  Suppose the system is in the state as shown in Fig.~\ref{fig:MDSqueue_working_a}, wherein the jobs $A_2$, $A_1$, $B_1$ and $B_2$ are being served by the four servers, and there are three more batches waiting in the buffer. Suppose server $1$ completes servicing job $A_2$ (Fig.~\ref{fig:MDSqueue_working_b}). This server is now free to serve any of the $6$ jobs waiting in the buffer. Since jobs are processed only in order, it begins serving job $C_1$ (assignment of $C_2$ would also have been valid). Next, suppose server $1$ completes $C_1$ before any other servers complete their tasks (Fig.~\ref{fig:MDSqueue_working_c}). In this case, since server $1$ has already served a job of batch $C$, it is not allowed to service $C_2$. However, it can service any job from the next batch $\{D_1,D_2\}$, and one of these two jobs is (arbitrarily) assigned to it. Finally, when one of the other servers completes its service, that server is assigned job $C_2$ (Fig.~\ref{fig:MDSqueue_working_d}).

\label{eg:MDSqueue_working}
\end{example}

\subsection{Other Applications} The MDS queue also arises in other applications that require diversity or error correction. For instance, consider a system with $n$ processors, with the arriving jobs comprising computational tasks. It is often the case that the processors are not completely reliable~\cite{borkar2005designing}, and may give incorrect outputs at random instances. In order to guarantee a correct output, a job may be processed at $k$ different servers, and the results aggregated (perhaps by majority rule) to obtain the final answer. Such a system results precisely in an MDS(n,k) queue (with some arrival and service-time distributions).
In general, queues where jobs require diversity, for purposes such as security, error-protection etc., may be modelled as an MDS queue.~\footnote{An analogy that the academic will relate to is that of reviewing papers. There are $n$ reviewers in total, and each paper must be reviewed by $k$ reviewers. This forms an MDS(n,k) queue. The values of $\lambda$ and $\mu$ considered should be such that $\frac{\lambda}{\mu}$ is close to the maximum throughput, modelling the fact that reviewers are generally busy.} Finally, even in the setting of distributed storage systems, the MDS queue need not be restricted to analysing Maximum-Distance-Separable codes alone, and can be used for any code that supports recovery of the files from `any $k$ out of the $n$' nodes. 

\subsection{Exact Analysis} An exact analysis of the MDS queue is hard. The difficulty arises from the special property of the MDS queue, that each of the $k$ jobs of a batch must be served by $k$ \textit{distinct} servers. Thus, a Markov-chain representation of this queue is required to have each state encapsulating not only the number of batches or jobs in the queue, but also the configuration of each batch in the queue, i.e., the number of jobs of each batch currently being processed, the number completed processing, and the number still waiting in the buffer. Thus, when there are $b$ batches in the system, the system can have $\BigOmega{b^k}$ possible configurations. Since the number of batches $b$ in the system can take any value in $\{0,1,2,\ldots\}$, this leads to a Markov chain which has a state space that has infinite states in at least $k$ dimensions. Furthermore, the transitions along different dimensions are tightly coupled. This makes the Markov chain hard to analyse, and in this paper, we provide scheduling policies for the MDS queue that lower/upper bound the exact MDS queue.

\section{Related Literature}\label{sec:literature}
\subsection{Queueing-theoretic analysis of coded systems}
The study of latency of coded systems was initiated by Huang et al. in~\cite{huang2012codes}, which we build upon in this paper. In particular, in~\cite{huang2012codes}, a `block-one-scheduling' policy was presented, that provides a lower bound on the performance of the MDS queue with $k=2$. This policy is a special case of the MDS-Reservation(t) policies presented in this paper, and corresponds to the case when t$\ =1$. While the block-one-scheduling policy was analysed only for $k=2$ in~\cite{huang2012codes}, the analysis in this paper applies to all values of $(n,k)$, and recovers the corresponding results of~\cite{huang2012codes} as a special case. In addition, the MDS-Reservation(t) policies of the present paper, when t$\ > 1$, provide significantly tighter bounds to the performance of the MDS queue (as can be seen in Fig.~\ref{fig:preview}). The present paper also provides upper bounds to the performance of the MDS queue.

The blocking probability of such systems in the absence of a buffer was previously studied in~\cite{ferner2012toward}. 

\subsection{Fork-join queues for parallel processing}
A class of queues that are closely related to the MDS queue is the class of \textit{fork-join queues}~\cite{baccelli1989fork}, and in particular, fork-join queues with variable subtasks~\cite{varki2008arif}. The classical setup of fork-join queues assumes each batch to consist of $n$ jobs, while the setup of fork-join queues with variable subtasks assumes $k$ jobs per batch for some parameter $k\leq n$. However, under a fork-join queue (with variable subtasks), each job must be served by a particular \textit{pre-specified} server, while under an MDS queue, the $k$ jobs of a batch may be processed at any \textit{arbitrary} set of $k$ servers and this choice is governed by the scheduling policy.

\subsection{Redundant Requests}\label{sec:literature_redundant_requests}
In a system that employs a $(n,k)$ erasure code, the latency of serving the requests can potentially be reduced by sending the requests redundantly to more than $k$ servers. The request is deemed served when it is served in any one of these ways. Following this, the other copies of this request may be removed from the system. A theoretical analysis when redundant requests help is initiated in~\cite{shah2013redundant} under the MDS Queue model (and also under a distributed counterpart). It is shown in~\cite{shah2013redundant} that, surprisingly, for any arbitrary arrival process, sending each request (redundantly) to all $n$ servers results in the smallest average latency (as compared to any other redundant-requesting policy) when the service time distribution is memoryless or heavier. Bounds on the average latency when the requests are sent (redundantly) to \textit{all} $n$ servers are derived in an independent work~\cite{joshi2012coding}. Approximations and empirical evaluations are performed in~\cite{dean2013tail,liang2013fast,vulimiri2012more,ananthanarayanan2012let}.

\section{Our Approach and Notation for Latency Analysis of The MDS Queue}\label{sec:approach}
For each of the scheduling policies presented in this paper (that lower/upper bound the MDS queue), we represent the respective resulting queues as continuous time Markov chains. We show that these Markov chains belong to a class of processes known as \textit{Quasi-Birth-Death (QBD)} processes (described below), and obtain their steady-state distribution by exploiting the properties of QBD processes. This is then employed to compute other metrics such as the average latency, system occupancy, etc.

Throughout the paper, we shall refer to the entire setup described in Section~\ref{sec:definition} as the `queue' or the `system'. We shall say that a batch is waiting (in the buffer) if at-least one of its jobs is still waiting in the buffer (i.e., has not begun service). We shall use the term ``\textit{$i\supth$ waiting batch}'' to refer to the batch that was the $i\supth$ earliest to arrive, among all batches currently waiting in the buffer. For example, in the system in the state depicted in Fig.~\ref{fig:MDSqueue_working_a}, there are three waiting batches: $\{C_1,C_2\}$, $\{D_1,D_2\}$ and $\{E_1,E_2\}$ are the first, second and third waiting batches respectively.

We shall frequently refer to an MDS queue as MDS(n,k) queue, and assume $[\lambda,\mu]$ to be some fixed (known) values. The system will always be assumed to begin in a state where there are no jobs in the system. Since the arrival and service time distributions have valid probability density functions,  we shall assume that no two events occur at exactly the same time. We shall use the notation $a^+$ to denote $\max(a,0)$.

\textit{Review of Quasi-Birth-Death (QBD) processes}: Consider a continuous-time Markov process on the states $\{0,1,2,\ldots\}$, with transition rate $\lambda_0$ from state $0$ to $1$, $\lambda$ from state $i$ to $(i+1)$ for all $i\geq 1$, $\mu_0$ from state $1$ to $0$, and $\mu$ from state $(i+1)$ to $i$ for all $i \geq 1$. This is a birth-death process. A QBD process is a generalization of such a birth-death process, wherein, each state $i$ of the birth-death process is replaced by a set of states. The states in the first set (corresponding to $i=0$ in the birth-death process) is called the set of \textit{boundary} states, whose behaviour is permitted to differ from that of the remaining states. The remaining sets of states are called the \textit{levels}, and the levels are identical to each other (recall that all states $i \geq 1$ in the birth-death process are identical). The Markov chain may have transitions only within a level or the boundary, between adjacent levels, and between the boundary and the first level. The transition probability matrix of a QBD process is thus of the form
\[
\begin{bmatrix}
 		B_1 & B_2 & 0 & 0  & \cdots ~\\
 		B_0 & A_1 & A_2 & 0  & \cdots ~\\
 		0 & A_0 & A_1 & A_2 & \cdots~ \\
 		0 & 0 & A_0 & A_1 & \cdots ~\\
 		0 & 0 & 0 & A_0 &  \cdots~ \\
 		\vdots & \vdots & \vdots & \vdots  &\ddots ~\\
     \end{bmatrix}.
\label{eq:QBD_matrix}\]
Here, the matrices $B_0, ~B_1, ~B_2, ~A_0, ~A_1$ and $A_2$ represent transitions entering the boundary from the first level, within the boundary, exiting the boundary to the first level, entering a level from the next level, within a level, and exiting a level to the next level respectively. If the number of boundary states is $q_b$, and if the number of states in each level is $q_\ell$, then the matrices $B_0$, $B_1$ and $B_2$ have dimensions $(q_\ell \times q_b)$, $(q_b \times q_b)$ and $(q_b \times q_\ell)$ respectively, and each of $A_0$, $A_1$ and $A_2$ have dimensions $(q_\ell \times q_\ell)$. The birth-death process described above is a special case with $q_b = q_\ell = 1$ and $B_0=\mu_0$, $B_1=0$, $B_2 =\lambda_0$, $A_0=\mu$, $A_1=0$, $A_2=\lambda$. Figures~\ref{fig:MDS_Reservation0_n4k2_statetransitiondiag},~\ref{fig:MDS_Reservation1_n4k2_statetransitiondiag} and~\ref{fig:MkMn0_n4k2_statetransitiondiag} in the sequel also present examples of QBD processes.
 
QBD processes are very well understood~\cite{feldman2010applied}, and their stationary distribution is fairly easy to compute. In this paper, we employ the \textit{SMCSolver} software package~\cite{bini2006structured} for this purpose. In the next two sections, we present scheduling policies which lower and upper bound the performance of the MDS queue, and show that the resulting queues can be represented as QBD processes. This representation makes them them easy to analyse, and this is exploited subsequently in the analysis presented in Section~\ref{sec:analysis}.

\section{Lower Bounds: MDS-Reservation(\lowercase{$t$}) Queues}\label{sec:lower}
This section presents a class of scheduling policies (and resulting queues), which we call the MDS-Reservation(t) scheduling policies (and MDS-Reservation(t) queues), whose performance lower bounds the performance of the MDS queue. This class of scheduling policies are indexed by a parameter `t': a higher value of t leads to a better performance and a tighter lower bound to the MDS queue, but on the downside, requires maintenance of a larger state and is also more complex to analyse.

The MDS-Reservation(t) scheduling policy, in a nutshell, is as follows: 
\begin{center}
\begin{minipage}{0.95\linewidth}  
\vspace{-.1cm}
 ``\textit{apply the MDS scheduling policy, but with an additional restriction that for any $i \in \{t+1, t+2,\ldots\}$, the $i\supth$ waiting batch is allowed to move forward in the buffer only when all $k$ of its jobs can move forward together.}''
\end{minipage}
\end{center}

We first describe in detail the special cases of t$=0$ and t$=1$, before moving on to the scheduling policy for a general t.

\subsection{MDS-Reservation(0)}
\subsubsection{Scheduling policy} 
The MDS-Reservation(0) scheduling policy is rather simple: the batch at the head of the buffer may start service only when $k$ or more servers are idle. The policy is described formally in Algorithm~\ref{alg:Reservation0}.
\begin{algorithm}[H]
\caption{MDS-Reservation(0) Scheduling Policy}
\begin{algorithmic}
\algostate \textbf{On} arrival of a batch
\algostate \algoindent \textbf{If} {number of idle servers $<k$}
\algostate \algoindent \algoindent append new batch at the end of buffer
\algostate \algoindent \textbf{Else}
\algostate \algoindent \algoindent assign $k$ jobs of the batch to any $k$ idle servers
\algostate \textbf{On} departure from server
\algostate \algoindent \textbf{If} {(number of idle servers $\!\geq\! k$) and (buffer is non-empty)}
\algostate \algoindent \algoindent  assign $k$ jobs of the first waiting batch to any $k$ idle servers
\end{algorithmic}
\label{alg:Reservation0}
\end{algorithm}

\begin{example}
Consider the MDS(n=4,k=2) queue in the state depicted in Fig.~\ref{fig:MDSqueue_working_a}. Suppose the server $2$ completes processing job $A_1$ (Fig.~\ref{fig:Reservation0_workinga}). Upon this event, the MDS scheduling policy would have allowed server $2$ to take up execution of either $C_1$ or $C_2$. However, this is not permitted under MDS-Reservation(0), and this server remains idle until a total of at least $k=2$ servers become idle. Now suppose the third server completes execution of $B_1$ (Fig.~\ref{fig:Reservation0_workingb}). At this point, there are sufficiently many idle servers to accommodate all $k=2$ jobs of the batch $\{C_1,C_2\}$, and hence jobs $C_1$ and $C_2$ are assigned to servers $2$ and $3$.
\begin{figure}[t]
\centering
\subfloat[]{
\includegraphics[width=.16\textwidth]{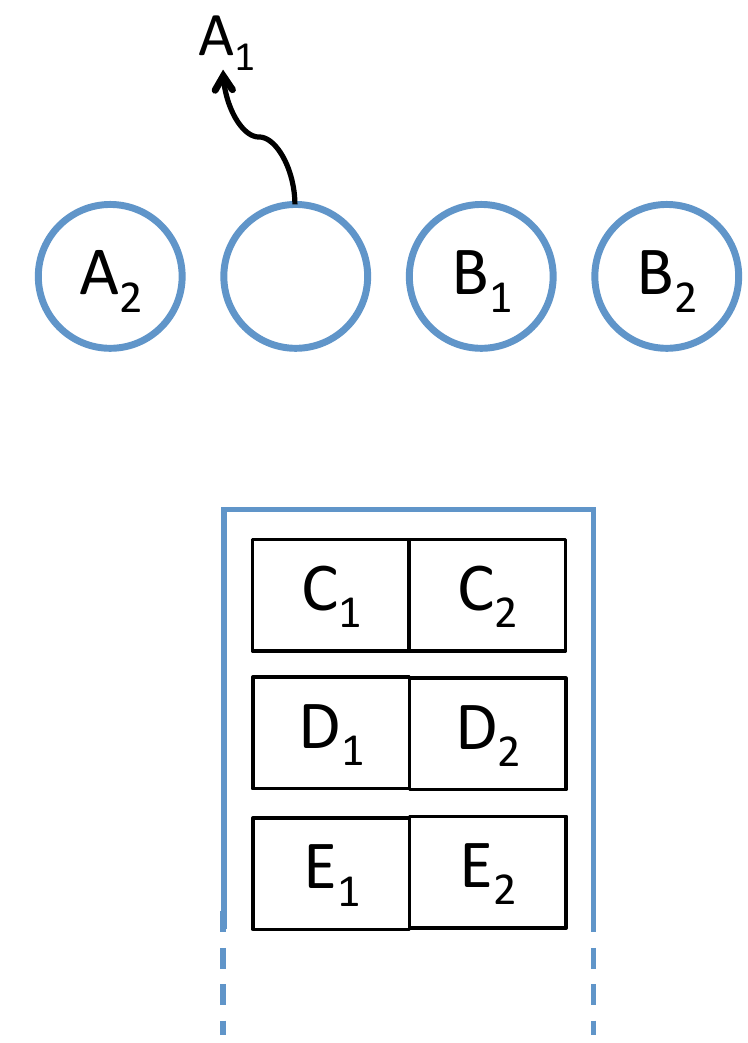}
\label{fig:Reservation0_workinga}
}~
\subfloat[]{
\includegraphics[width=.16\textwidth]{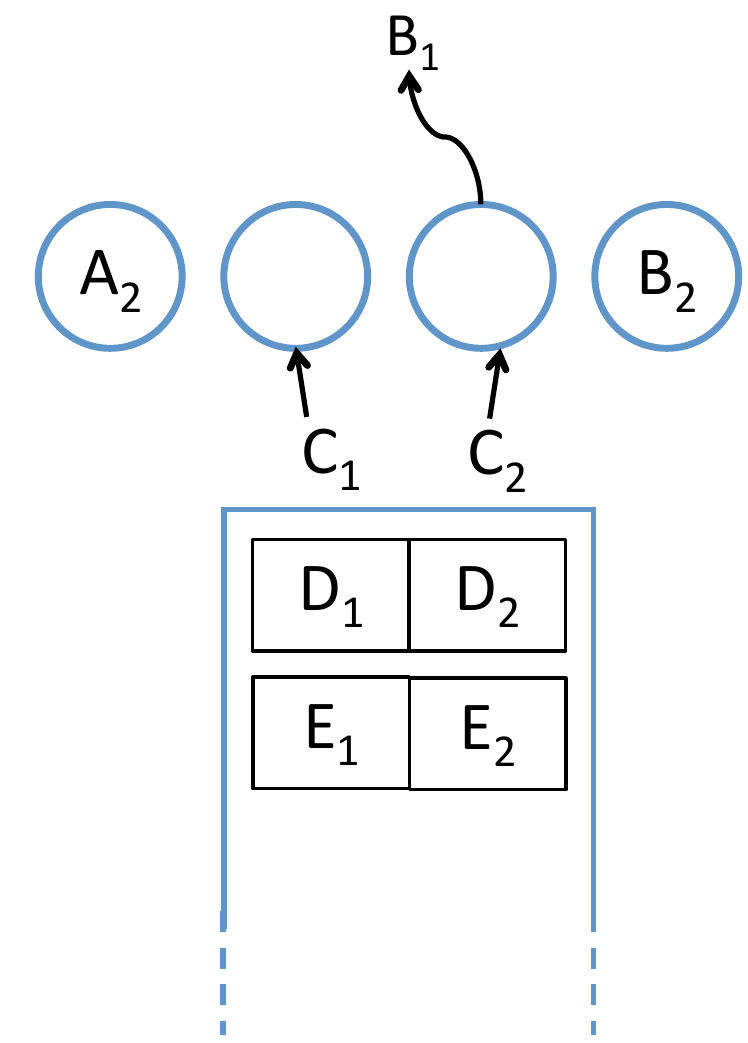}
\label{fig:Reservation0_workingb}
}
\caption{An illustration of the MDS-Reservation(0) scheduling policy for a system with parameters $(n=4,k=2)$. This policy prohibits the servers to process jobs from a batch unless there are $k$ idle servers that can process all $k$ jobs of that batch. As shown in the figure, server $1$ is barred from processing $\{C_1,C_2\}$ in (a), but is subsequently allowed to do so when another server also becomes idle in (b).}
\label{fig:Reservation0_working}
\end{figure}
\end{example}


We note that the MDS-Reservation(0) queue, when $n=k$, is identical to a split-merge queue~\cite{harrison2003queueing}.

\subsubsection{Analysis}
Observe that under the specific scheduling policy of MDS-Reservation(0), a batch that is waiting in the buffer must necessarily have all its $k$ jobs in the buffer, and furthermore, these $k$ jobs go into the servers at the same time.

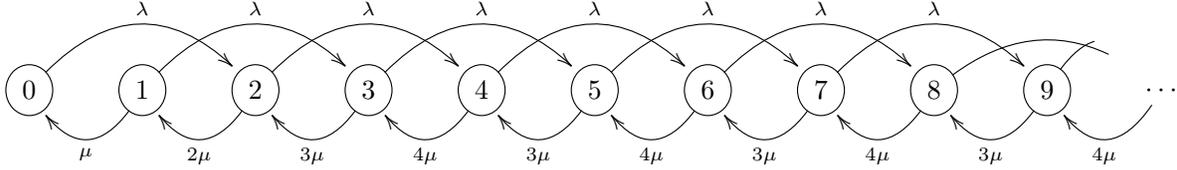
\begin{figure}[t!]
\centering
\medmuskip=0\medmuskip
\resizebox{.99\textwidth}{!}{
\centering
\hspace{1cm}\xymatrix{
*++[o][F]{0} \ar@/^2pc/[rr]^{\lambda} &
*++[o][F]{1} \ar@/^2pc/[rr]^{\lambda} \ar@/^1.5pc/[l]^{\mu}&
*++[o][F]{2} \ar@/^2pc/[rr]^{\lambda}
  \ar@/^1.5pc/[l]^{2\mu}&
*++[o][F]{3} \ar@/^2pc/[rr]^{\lambda} \ar@/^1.5pc/[l]^{3\mu}&
*++[o][F]{4} \ar@/^2pc/[rr]^{\lambda}
  \ar@/^1.5pc/[l]^{4\mu}&
*++[o][F]{5} \ar@/^2pc/[rr]^{\lambda} \ar@/^1.5pc/[l]^{3\mu}&
*++[o][F]{6} \ar@/^2pc/[rr]^{\lambda}
  \ar@/^1.5pc/[l]^{4\mu}&
*++[o][F]{7} \ar@/^2pc/[rr]^{\lambda} \ar@/^1.5pc/[l]^{3\mu}&
*++[o][F]{8} \ar@{halfTwenty}@/^1.52pc/[rr]
  \ar@/^1.5pc/[l]^{4\mu}&
*++[o][F]{9} \ar@{halfForty}@/^1.52pc/[r] \ar@/^1.5pc/[l]^{3\mu}&
\cdots \ar@/^1.5pc/[l]^{4\mu} &
}
}
\caption{State transition diagram of the MDS-Reservation(0) queue for $n=4$ and $k=2$. The notation at any state is the number of jobs $m$ in the system in that state. The set of boundary states are $\{0, 1, 2\}$, and the levels are pairs of states $\{3,4\}$, $\{5,6\}$, $\{7,8\}$, etc. The transition matrix is of the form~\eqref{eq:QBD_matrix} with $B_0 = [0~~~0~~~3\mu~;~~0~~~0~~~0]$, $B_1=[-\lambda~~~0~~~\lambda~;~~\mu~~~-(\mu+\lambda)~~~0;~~0~~~2\mu~~~-(2\mu+\lambda)]$, $B_2=[0~~~0~;~~\lambda~~~0~;~~~0~~~\lambda]$, $A_0 = [0~~~3\mu~;~~0~~~0]$, $A_1=[-(3\mu+\lambda)~~~0~;~~4\mu~~~-(4\mu+\lambda)]$, $A_2=[\lambda~~~0~;~~0~~~\lambda]$. }
\label{fig:MDS_Reservation0_n4k2_statetransitiondiag}
\end{figure}

We now describe the Markovian representation of the MDS-Reservation(0) queue. We show that it suffices to keep track of only the total number of jobs $m$ in the entire system.
\begin{theorem}
A Markovian representation of the MDS-Reservation(0) queue has a state space $\{0,1,\ldots,\infty\}$, and any state $m \in \{0,1,\ldots,\infty\}$ has transitions to: (i) state $(m + k)$ at rate $\lambda$, (ii) if $m \leq n$ then to state $(m-1)$ at rate $m\mu$, and (iii) if $m>n$ then to state $(m-1)$ at rate $(n-(n-m) \modulo k))\mu$.
The MDS-Reservation(0) queue is thus a QBD process, with boundary states $\{0,1,\ldots,n-k\}$, and levels $m\in \{n-k+1 + jk,\ldots,n+jk\}$ for $j=\{0,1,\ldots,\infty\}$. 
\label{thm:MDS_Reservation0_transitions}
\end{theorem}
The state transition diagram of the MDS-Reservation(0) queue for $(n=4,k=2)$ is depicted in Fig.~\ref{fig:MDS_Reservation0_n4k2_statetransitiondiag}.

\newdir{halfFive}{!/5pt/@{ }*:(0,0)@^{ }*:(0,0)@_{ }}
\newdir{halfTen}{!/10pt/@{ }*:(0,0)@^{ }*:(0,0)@_{ }}
\newdir{halfTwenty}{!/20pt/@{ }*:(0,0)@^{ }*:(0,0)@_{ }}
\newdir{halfThirty}{!/30pt/@{ }*:(0,0)@^{ }*:(0,0)@_{ }}
\newdir{halfForty}{!/40pt/@{ }*:(0,0)@^{ }*:(0,0)@_{ }}
\newdir{halfFifty}{!/50pt/@{ }*:(0,0)@^{ }*:(0,0)@_{ }}
\newdir{halfSixty}{!/60pt/@{ }*:(0,0)@^{ }*:(0,0)@_{ }}
\newdir{halfSeventy}{!/70pt/@{ }*:(0,0)@^{ }*:(0,0)@_{ }}
\newdir{halfHundred}{!/100pt/@{ }*:(0,0)@^{ }*:(0,0)@_{ }}

Theorem~\ref{thm:MDS_Reservation0_transitions} shows that the MDS-Reservation(0) queue is a QBD process, allowing us to employ the SMC solver to obtain its steady-state distribution. Alternatively, the MDS-Reservation(0) queue is simple enough to analyse directly as well. To this end, let $y(m)$ denote the number of jobs being served when the Markov chain is in state $m$. From the description above, this function can be written as:
\begin{align*}
y(m) = \begin{cases} m, & \text{if} ~~0 \leq i \leq n
			\\ n-((n-m) \modulo k), & \text{if} ~~m >n~.
\end{cases}
\end{align*}
Let $\mathbf{\pi} = [\pi_0~\pi_1~\pi_2~\cdots]$ denote the steady-state distribution of this chain. The global balance equation for the cut between states $(m-1)$ and $m$ gives:
\beq
\pi_{m} = \frac{\lambda}{y(m)\mu}\left(\sum_{j=(m-k)^+}^{m-1}{\pi_{j}}\right)~~~\forall~m>0. \eeq
Using these recurrence equations, for any given $(n,k)$, the distribution $\mathbf{\pi}$ of the number of jobs in steady-state can be computed easily.


\subsection{MDS-Reservation(1)}
\subsubsection{Scheduling policy} 
The MDS-Reservation(0) scheduling policy discussed above allows the batches in the buffer to move ahead only when all $k$ jobs in the batch can move together. The MDS-Reservation(1) scheduling policy relaxes this restriction for (only) the job at the head of the buffer. This is formalized in Algorithm~\ref{alg:Reservation1}.
\begin{algorithm}[H]
\caption{MDS-Reservation(1) Scheduling Policy}
\begin{algorithmic}
\algostate \textbf{On} arrival of a batch
\algostate \algoindent \textbf{If} {buffer is empty}
\algostate \algoindent \algoindent assign one job each from new batch to idle servers
\algostate \algoindent append remaining jobs of batch to the end of the buffer
\algostate \textbf{On} departure from server (say, server $s$):
\algostate \algoindent \textbf{If} {buffer is non-empty and no job from first waiting batch has been served by $s$}
\algostate \algoindent \algoindent assign a job from first waiting batch to $s$
\algostate \algoindent \algoindent \textbf{If} {first waiting batch had only one job in buffer \& there exists another waiting batch}
\algostate \algoindent \algoindent \algoindent to every remaining idle server, assign a job from second waiting batch
\end{algorithmic}
\label{alg:Reservation1}
\end{algorithm}

\begin{example}
\begin{figure}[t]
\centering
\subfloat[]{
\includegraphics[width=.16\textwidth]{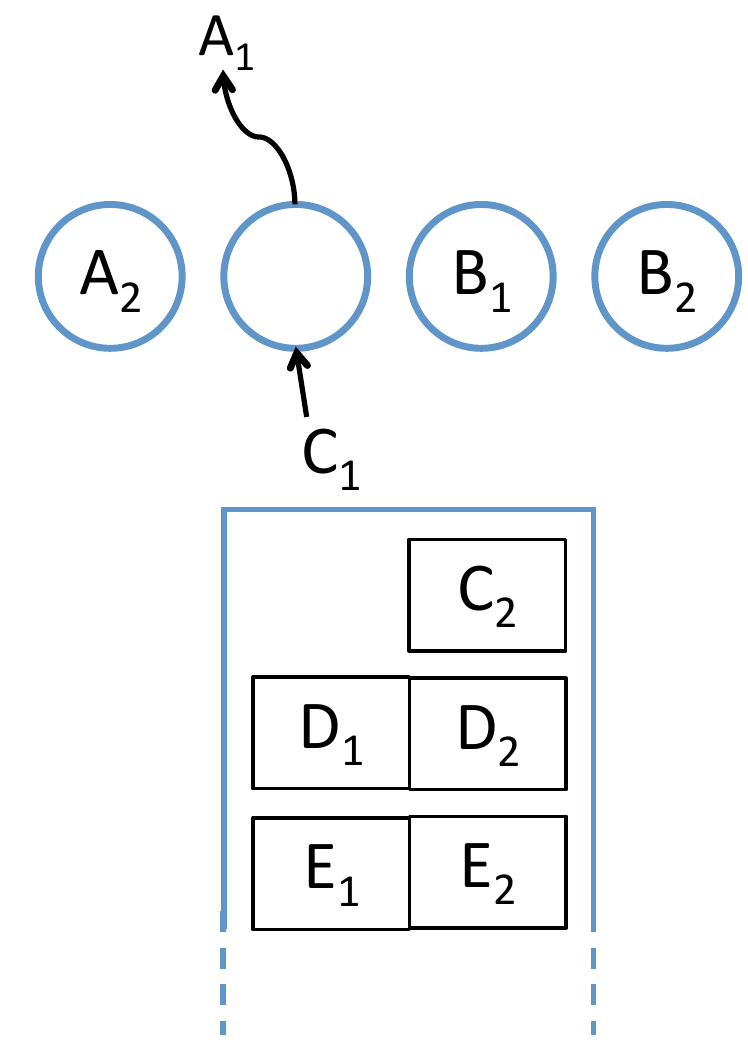}
\label{fig:Reservation1_workinga}
}~
\subfloat[]{
\includegraphics[width=.16\textwidth]{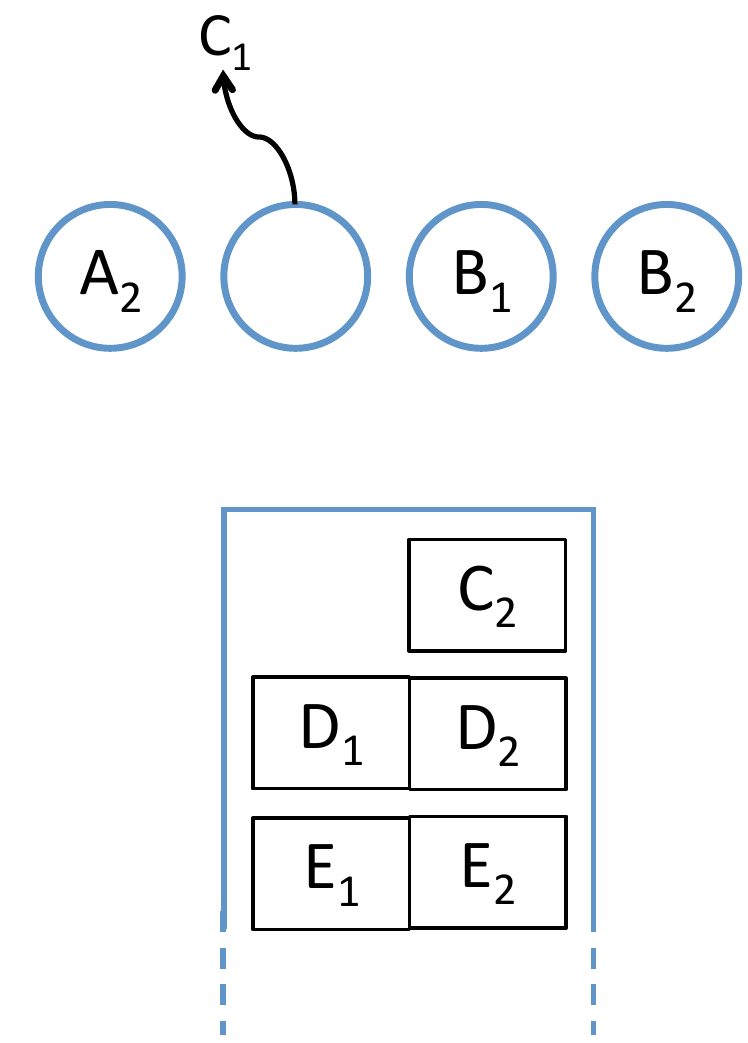}
\label{fig:Reservation1_workingb}
}~
\subfloat[]{
\includegraphics[width=.16\textwidth]{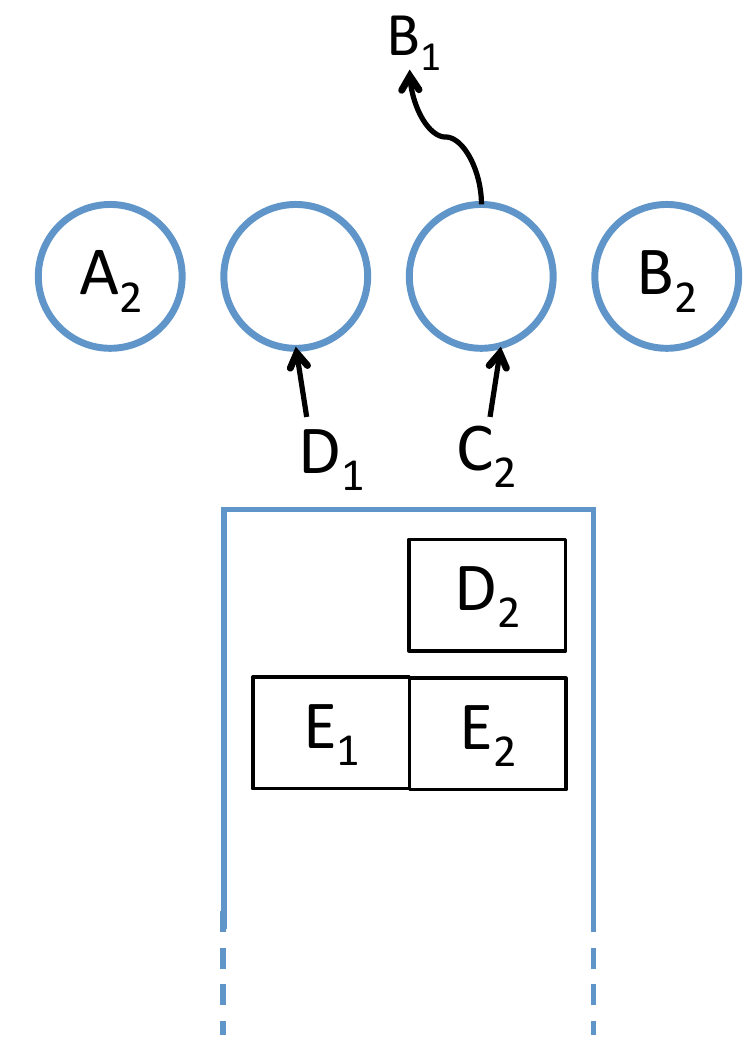}
\label{fig:Reservation1_workingc}
}
\caption{An illustration of the MDS-Reservation(1) scheduling policy, for a system with parameters $(n=4,k=2)$. As shown in the figure, this policy prohibits the servers from processing jobs of the second or later batches (e.g., $\{D_1,D_2\}$ and $E_1,E_2$ in (b)), until they move to the top of the buffer (e.g., $\{D_1,D_2\}$ in (c)).}
\label{fig:Reservation1_working}
\end{figure}
Consider the MDS(n=4,k=2) queue in the state depicted in Fig.~\ref{fig:MDSqueue_working_a}. Suppose server $2$ completes processing job $A_1$ (Fig.~\ref{fig:Reservation1_workinga}). Under MDS-Reservation(1), server $2$ now begins service of job $C_1$ (which is allowed by MDS, but was prohibited under MDS-Reservation(0)). Now suppose that server $2$ finishes this service before any other server (Fig.~\ref{fig:Reservation1_workingb}). In this situation, since server $2$ has already processed one job from batch $\{C_1,C_2\}$, it is not allowed to process $C_2$. However, there exists another batch $\{D_1,D_2\}$ in the buffer such that none of the jobs in this batch have been processed by the idle server $2$. While the MDS scheduling policy would have allowed server $2$ to start processing $D_1$ or $D_2$, this is not permitted under MDS-Reservation(1), and the second server remains idle. Now, if server $3$ completes service (Fig.~\ref{fig:Reservation1_workingc}), then $C_2$ is assigned to server $3$, allowing batch $\{D_1,D_2\}$ to move up as the first batch. This now permits server $2$ to begin service of job $D_1$.
\end{example}

The MDS-Reservation(1) scheduling policy is identical to the block-one-Scheduling policy proposed in~\cite{huang2012codes}. While this scheme was analysed in~\cite{huang2012codes} only for the case of $k=2$, in this paper, we present a more general analysis that holds for all values of the parameter $k$.

\subsubsection{Analysis}
The following theorem describes the Markovian representation of the MDS-Reservation(1) queue. Each state in this representation is defined by two quantities: (i) the total number of jobs $m$ in the system, and (ii) the number of jobs $w_1$ of the first waiting batch, that are still in the buffer. 

\begin{figure}[t!]
\centering
\resizebox{.99\textwidth}{!}{
\begin{tikzpicture}[->,>=stealth',shorten >=1pt,auto,node distance=1.8cm, semithick]
\tikzstyle{every state}=[fill=white,draw=black,thick,text=black,scale=1]
\node[state](S000){$0,0$};
\node[state](S001)[right of=S000]{$0,1$};
\node[state](S002)[right of=S001]{$0,2$};
\node[state](S003)[right of=S002]{$0,3$};
\node[state](S004)[right of=S003]{$0,4$};
\node[state](S101)[right of=S004]{$1,5$};
\node[state](S102)[right of=S101]{$2,6$};
\node[state](S201)[right of=S102]{$1,7$};
\node[state](S202)[right of=S201]{$2,8$};
\node[state](S301)[right of=S202]{$1,9$};
\node[state](S302)[right of=S301]{$2,10$};

\node[state](S111)[above of=S004]{$1,4$};
\node[state](S211)[above of=S102]{$1,6$};
\node[state](S311)[above of=S202]{$1,8$};
\node[state](S1_10)[above of=S302]{$1,10$};
\node[](dots1)[right of=S1_10]{$\cdots$};
\node[](dots2)[right of=S302]{$\cdots$};

\path (S000) edge [bend right=50] node[below] {$\lambda$} (S002);
\path (S001) edge [bend right=50] node[below] {$\lambda$} (S003);
\path (S002) edge [bend right=50] node[below] {$\lambda$} (S004);
\path (S003) edge [bend right=50] node[below] {$\lambda$} (S101);
\path (S004) edge [bend right=50] node[below] {$\lambda$} (S102);
\path (S101) edge [bend right=50] node[below] {$\lambda$} (S201);
\path (S102) edge [bend right=50] node[below] {$\lambda$} (S202);
\path (S201) edge [bend right=50] node[below] {$\lambda$} (S301);
\path (S202) edge [bend right=50] node[below] {$\lambda$} (S302);
\path (S311) edge [bend left=20] node[above] {$\lambda$} (S1_10);
\path (S111) edge [bend left=20] node[above] {$\lambda$} (S211);
\path (S211) edge [bend left=20] node[above] {$\lambda$} (S311);

\path (S001) edge [bend right] node[above] {$\mu$} (S000);
\path (S002) edge [bend right] node[above] {$2\mu$} (S001);
\path (S003) edge [bend right] node[above] {$3\mu$} (S002);
\path (S004) edge [bend right] node[above] {$4\mu$} (S003);
\path (S101) edge [bend right] node[above] {$3\mu$} (S004);
\path (S102) edge [bend right] node[above] {$4\mu$} (S101);
\path (S201) edge [bend right] node[above] {$3\mu$} (S102);
\path (S202) edge [bend right] node[above] {$4\mu$} (S201);
\path (S301) edge [bend right] node[above] {$3\mu$} (S202);
\path (S302) edge [bend right] node[above] {$4\mu$} (S301);

\path (S111) edge [bend right=20] node[above] {$3\mu$} (S003);
\path (S211) edge [bend right=20] node[above] {$3\mu$} (S101);
\path (S311) edge [bend right=20] node[above] {$3\mu$} (S201);
\path (S1_10) edge [bend right=20] node[above] {$3\mu$} (S301);

\path (S101) edge [bend right=20] node[right] {$\mu$} (S111);
\path (S201) edge [bend right=20] node[right] {$\mu$} (S211);
\path (S301) edge [bend right=20] node[right] {$\mu$} (S311);

\end{tikzpicture}
}
\caption{State transition diagram of the MDS-Reservation(1) queue for $n=4$ and $k=2$. The notation at any state is $(w_1,m)$. The subset of states that are never visited are not shown. The set of boundary states are $\{0, 1, 2,3,4\}\times\{0,1,2\}$, and the levels are sets $\{5,6\}\times\{0,1,2\}$, $\{7,8\}\times\{0,1,2\}$, etc.}
\label{fig:MDS_Reservation1_n4k2_statetransitiondiag}
\end{figure}
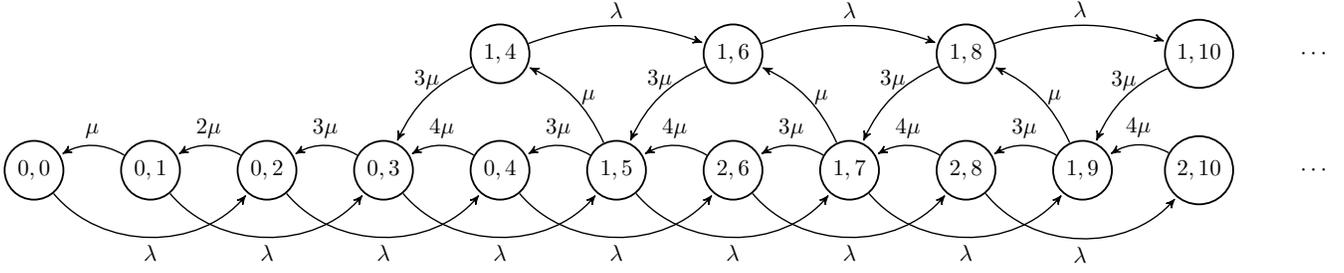

\begin{theorem}
The Markovian representation of the MDS-Reservation(1) queue has a state space $\{0,1,\ldots,k\} \times \{0,1,\ldots,\infty\}$. It is a QBD process with boundary states $\{0,\ldots,k\} \times \{0,\ldots,n\}$, and levels $\{0,\ldots,k\} \times \{n-k+1+jk,\ldots,n+jk\}$ for $j=\{1,2,\ldots,\infty\}$.
\label{thm:MDS_Reservation1_transitions}
\end{theorem}
The state transition diagram of the MDS-Reservation(1) queue for $(n=4,k=2)$ is depicted in Fig.~\ref{fig:MDS_Reservation1_n4k2_statetransitiondiag}.

Note that the state space $\{0,1,\ldots,k\}\times\{0,1,\ldots,\infty\}$ has several states that will never be visited during the execution of the Markov chain. For instance, the states $(w_1>0, m\leq n-k)$ never occur. This is because $w_1>0$ implies existence of some job waiting in the buffer, while $m \leq n-k$ implies that $k$ or more servers are idle. The latter condition implies that there exists at least one idle server that can process a job from the first waiting batch, and hence the value of $w_1$ must be smaller than that associated to that state, thus proving the impossibility of the system being in that state.

\subsection{MDS-Reservation(\lowercase{t}) for a general \lowercase{t}}
\subsubsection{Scheduling policy}
Algorithm~\ref{alg:Reservationt} formally describes the MDS-Reservation(t) scheduling policy.
\begin{algorithm}[h]
\caption{MDS-Reservation(t) Scheduling Policy}
\begin{algorithmic}
\algostate \textbf{On} arrival of a batch
\algostate \algoindent \textbf{If}  {buffer has strictly fewer than t batches}
\algostate \algoindent \algoindent  Assign jobs of new batch to idle servers
\algostate \algoindent Append remaining jobs of batch to end of buffer
\algostate \textbf{On} departure of job from a server (say, server $s$)
\algostate \algoindent {\medmuskip=0\medmuskip \thickmuskip=0\thickmuskip Find $\hat{i} =\min\{i\geq 1:$ $s$ has not served job of $i\supth$ waiting batch$\}$}
\algostate \algoindent Let $b_{t+1}$ be the $(t+1)\supth$ waiting batch (if any)
\algostate \algoindent \textbf{If} {$\hat{i}$ exists \& $\hat{i} \leq t$}
\algostate \algoindent \algoindent  Assign a job of $\hat{i}\supth$ waiting batch to $s$
\algostate \algoindent \algoindent \textbf{If} {$\hat{i}=1$ \& the first waiting batch had only one job in the buffer \& $b_{t+1}$ exists}
\algostate \algoindent \algoindent \algoindent To every remaining idle server, assign a job from batch $b_{t+1}$
\end{algorithmic}
\label{alg:Reservationt}
\end{algorithm}

The following example illustrates the MDS-Reservation(t) scheduling policy when t$=2$.
\begin{example}(t=2).\label{eg:Reservation2_working}
Consider the MDS(n=4,k=2) queue in the state depicted in Fig.~\ref{fig:MDSqueue_working_a}. Suppose the second server completes processing job $A_1$ (Fig.~\ref{fig:Reservation2_workinga}). Under the MDS-Reservation(2) scheduling policy, server $2$ now begins service of job $C_1$. Now suppose that server $2$ finishes this service as well, before any other server completes its respective service (Fig.~\ref{fig:Reservation2_workingb}). In this situation, while MDS-Reservation(1) would have mandated server $2$ to remain idle, MDS-Reservation(2) allows it to start processing a job from the next batch $\{D_1,D_2\}$. However, if the server also completes processing of this job before any other server (Fig.~\ref{fig:Reservation2_workingc}), then it is not allowed to take up a job of the third batch $\{E_1,E_2\}$. Now suppose server $3$ completes service (Fig.~\ref{fig:Reservation2_workingd}). Server $3$ can begin serving job $C_2$, thus clearing batch $\{C_1,C_2\}$ from the buffer, and moving the two remaining batches up in the buffer. Batch $\{E_1,E_2\}$ is now within the threshold of t$=2$, allowing it to be served by the idle server $2$.
\begin{figure}[t!]
\centering
\subfloat[]{
\includegraphics[width=.16\textwidth]{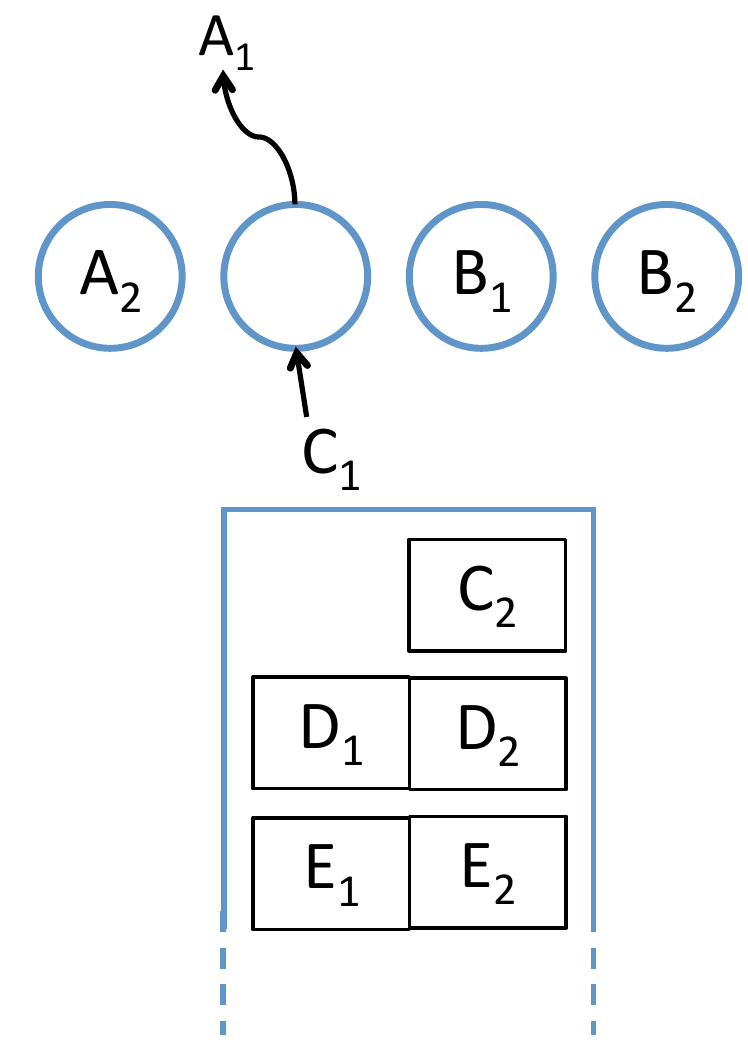}
\label{fig:Reservation2_workinga}
}~
\subfloat[]{
\includegraphics[width=.16\textwidth]{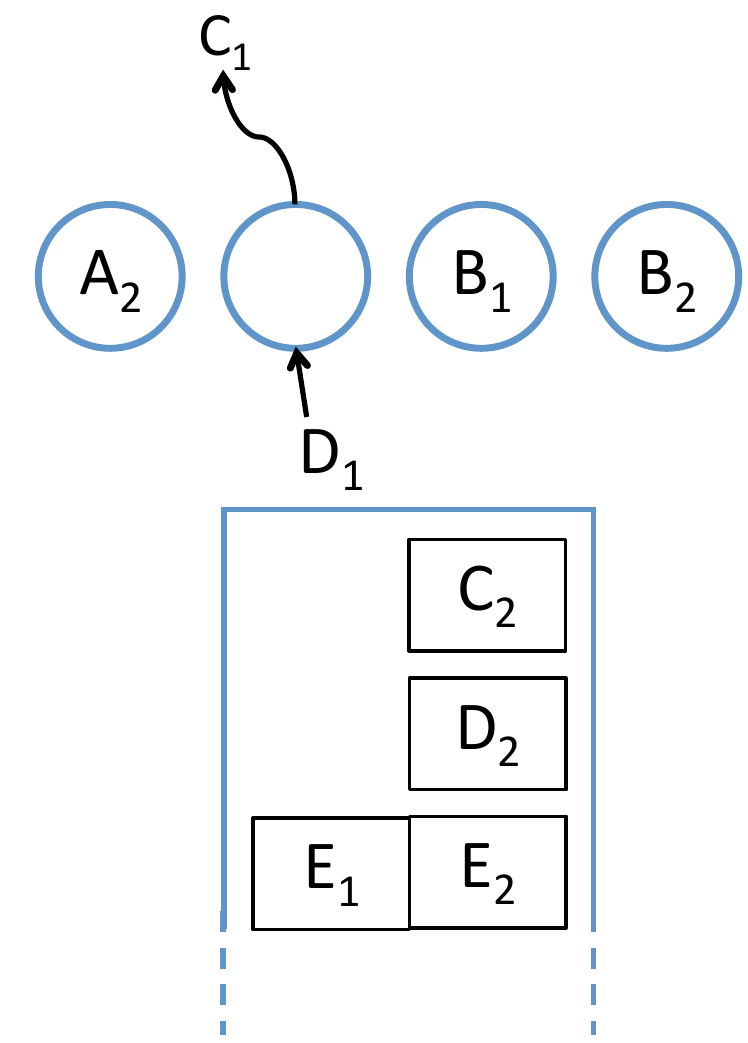}
\label{fig:Reservation2_workingb}
}~
\subfloat[]{
\includegraphics[trim=0 0 0 .05cm, clip=true, width=.16\textwidth]{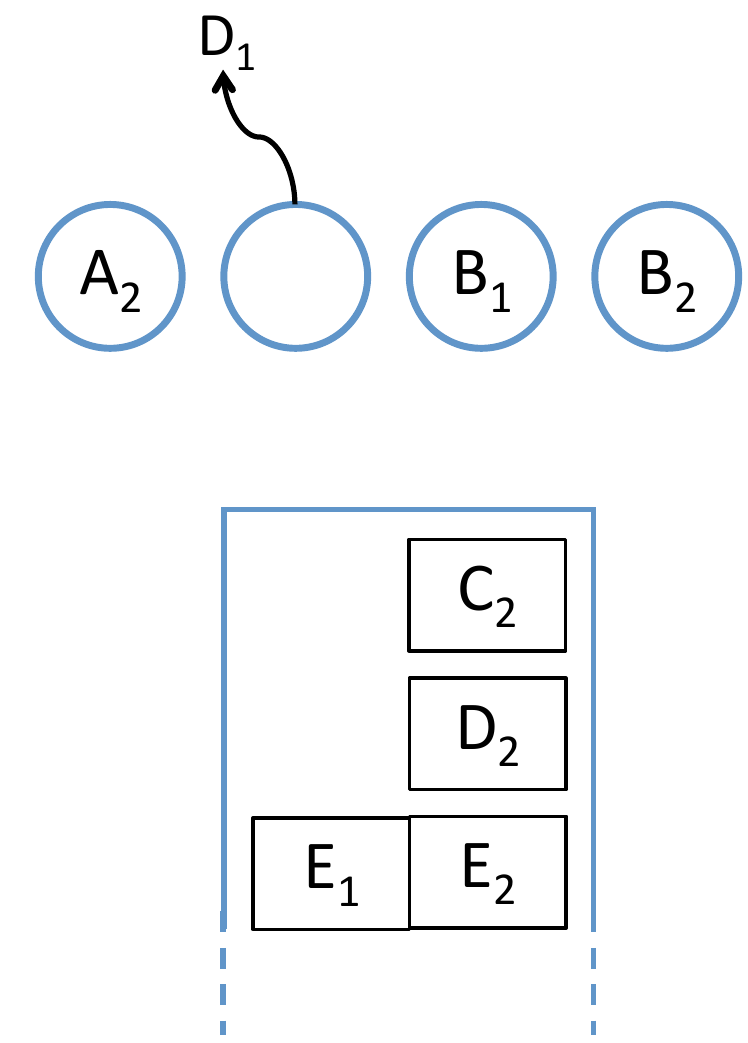}
\label{fig:Reservation2_workingc}
}~
\subfloat[]{
\includegraphics[trim=0 0 0 .05cm, clip=true, width=.16\textwidth]{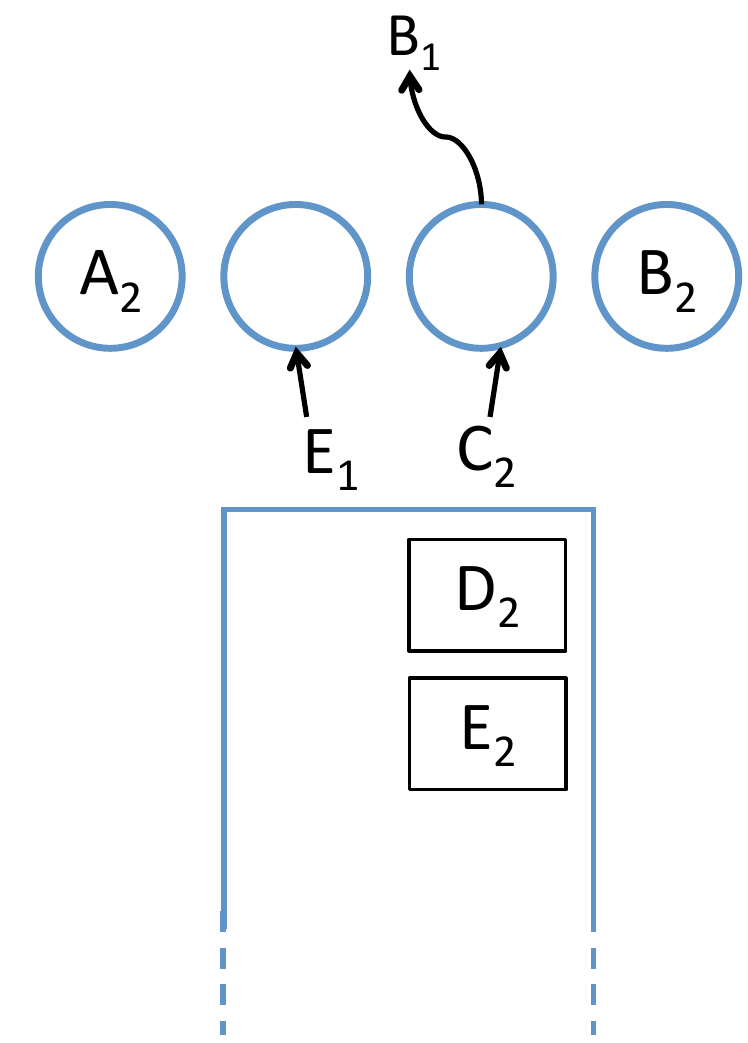}
\label{fig:Reservation2_workingd}
}~
\caption{An illustration of the working of the MDS-Reservation(2) scheduling policy, for a system with parameters $(n=4,k=2)$. As shown in the figure, this policy prohibits the servers from processing jobs of the third and later batches (e.g., batch $\{E_1,E_2\}$ in (c)), until they move higher in the buffer (e.g., as in (d)).}
\label{fig:Reservation2_working}
\end{figure}
\end{example}

\subsubsection{Analysis}
\begin{theorem}
The Markovian representation of the MDS-Reservation(t) queue has a state space $\{0,1,\ldots,k\}^t \times \{0,1,\ldots,\infty\}$. It is a QBD process with boundary states $\{0,\ldots,k\}^t \times \{0,\ldots,n-k+tk\}$, and levels $\{0,\ldots,k\}^t \times \{n-k+1+jk,\ldots,n+jk\}$ for $j=\{t,\,t+1,\ldots,\infty\}$.
\label{thm:MDS_Reservationt_transitions}
\end{theorem}
The proof of Theorem~\ref{thm:MDS_Reservationt_transitions} also describes how one can obtain the configuration of the entire system from only the number of jobs in the system, under the MDS-Reservation(t) scheduling policies.

One can see that the sequence of MDS-Reservation(t) queues, as t increases, becomes closer to the MDS queue. This results in tighter bounds, and also increased complexity of the transition diagrams. The limit of this sequence is the MDS queue itself.
\begin{theorem}
The MDS-Reservation(t) queue, when t$\ =\infty$, is precisely the MDS queue.
\label{thm:MDS_Reservation_infinity}
\end{theorem}

\section{Upper bounds: $M^k/M/n$(\lowercase{$t$}) Queues}\label{sec:upper}
In this section, we present a class of scheduling policies (and resulting queues), which we call the \MkMn(t) scheduling policies (and \MkMn(t) queues), whose performance upper bounds the performance of the MDS queue. The scheduling policies presented here relax the constraint of requiring the $k$ jobs in a batch to be processed by $k$ distinct servers. While the \MkMn(t) scheduling policies and the \MkMn(t) queues are not realizable in practice, they are presented here only to obtain upper bounds to the performance of the MDS queue.

The \MkMn(t) scheduling policy, in a nutshell, is as follows:
\begin{center}
\begin{minipage}{0.95\linewidth}  
\vspace{-.07cm}
 ``\textit{apply the MDS scheduling policy whenever there are t or fewer batches in the buffer; when there are more than t batches in the buffer, ignore the restriction requiring the $k$ jobs of a batch to be processed by distinct servers.}''
\end{minipage}
\end{center}

We first describe the \MkMn(0) queue in detail, before moving on to the general \MkMn(t) queues.

\subsection{\MkMn(0)}
\subsubsection{Scheduling policy} 
The \MkMn(0) scheduling policy operates by completely ignoring the restriction of assigning distinct servers to jobs of the same batch. This is described formally in Algorithm~\ref{alg:MkMn0}.
\begin{algorithm}[H]
\caption{\MkMn(0)}
\begin{algorithmic}
\algostate \textbf{On} arrival of a batch
\algostate \algoindent assign jobs of this batch to idle servers (if any)
\algostate \algoindent append remaining jobs at the end of the buffer
\algostate \textbf{On} departure from a server
\algostate \algoindent \textbf{If} {buffer is not empty}
\algostate \algoindent \algoindent assign a job from the first waiting batch to this server
\end{algorithmic}
\label{alg:MkMn0}
\end{algorithm}

Note that the \MkMn(0) queue is identical to the \MkMn~queue, i.e., an $M/M/n$ queue with batch arrivals.


The following example illustrates the working of the \MkMn(0) scheduling policy.
\begin{example}
\begin{figure}[t]
\centering
\subfloat[]{
\includegraphics[width=.16\textwidth]{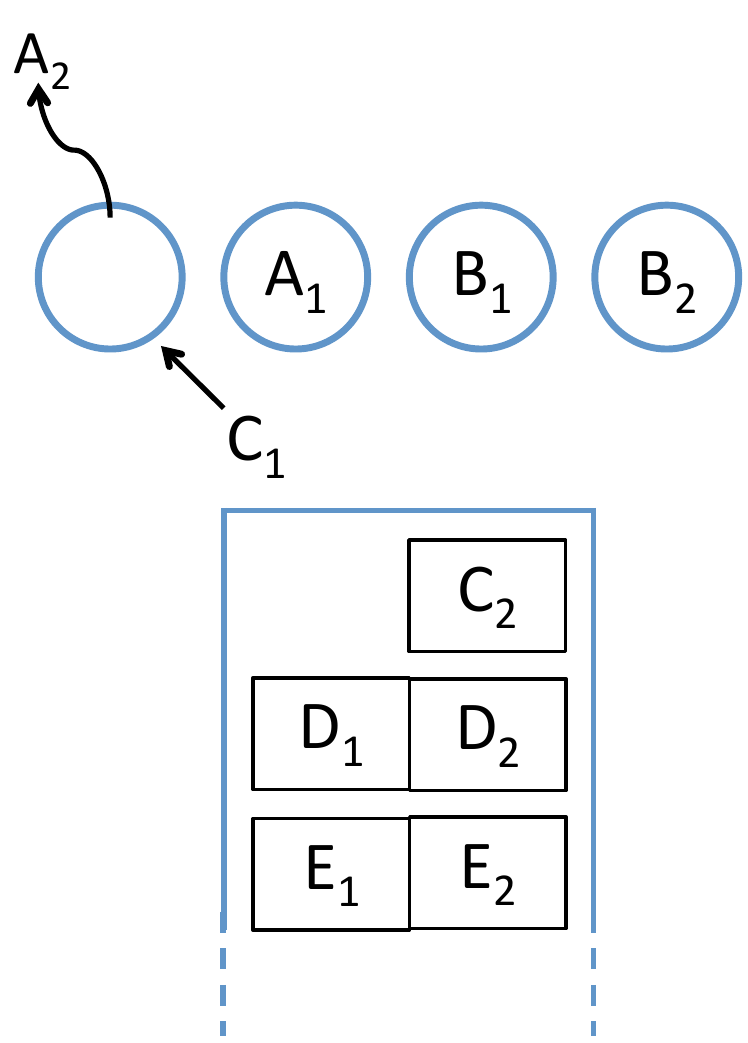}
\label{fig:MkMn0_workinga}
}~
\subfloat[]{
\includegraphics[width=.16\textwidth]{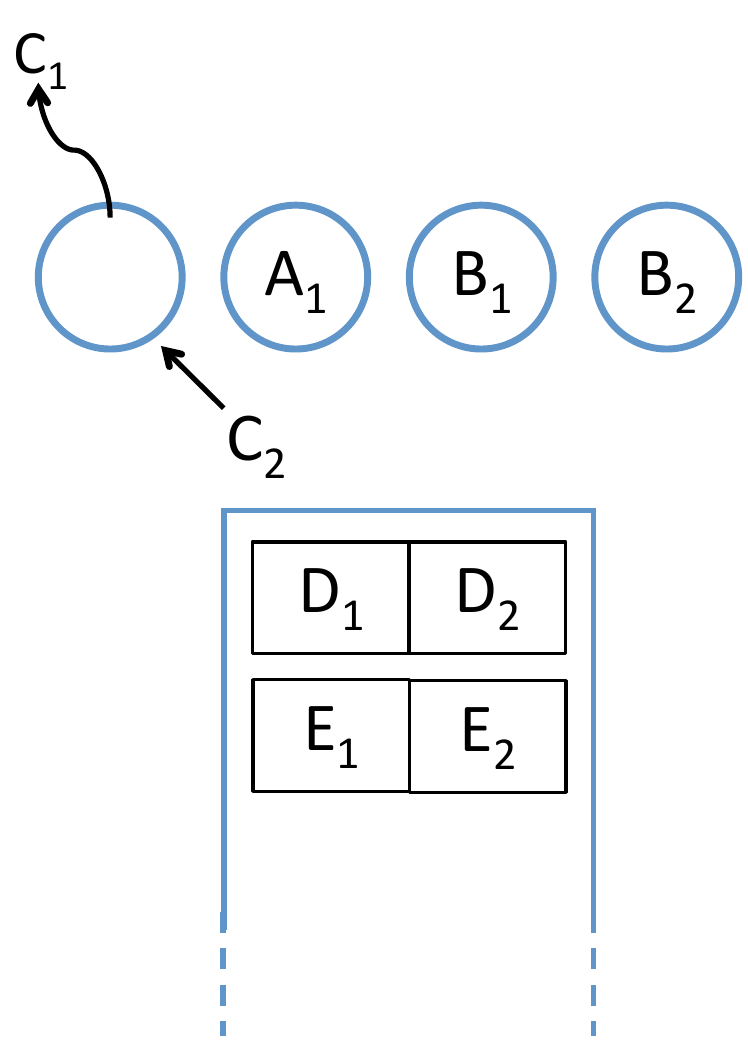}
\label{fig:MkMn0_workingb}
}
\caption{Illustration of the working of the \MkMn(0) scheduling policy. This policy allows a server to process more than one jobs of the same batch. As shown in the figure, server $1$ processes both $C_1$ and $C_2$.}
\label{fig:MkMn0_working}
\end{figure}
Consider the MDS(n=4,k=2) queue in the state depicted in Fig.~\ref{fig:MDSqueue_working_a}. Suppose the first server completes processing job $A_2$, as shown in Fig.~\ref{fig:MkMn0_workinga}. Under the \MkMn(0) scheduling policy, server $1$ now takes up job $C_1$. Next suppose server $1$ also finishes this task before any other server completes service (Fig.~\ref{fig:MkMn0_workingb}). In this situation, the MDS scheduling policy would prohibit job $C_2$ to be served by the first server. However, under the scheduling policy of \MkMn(0), we relax this restriction, and permit server $1$ to begin processing $C_2$.
\end{example}

\subsubsection{Analysis}
We now describe a Markovian representation of the \MkMn(0) scheduling policy, and show that it suffices to keep track of only the total number of jobs $m$ in the system.

\begin{theorem}
The Markovian representation of the \MkMn(0) queue has a state space $\{0,1,\ldots,\infty\}$, and any state $m \in \{0,1,\ldots,\infty\}$, has transitions (i) to state $(m+k)$ at rate $\lambda$, and (ii) if $m>0$, then to state $(m-1)$ at rate $\min(n,m)\mu$. It is a QBD process with boundary states $\{0,\ldots,k\} \times \{0,\ldots,n\}$, and levels $\{0,\ldots,k\} \times \{n-k+1+jk,\ldots,n+jk\}$ for $j=\{1,2,\ldots,\infty\}$.
\label{thm:MDS_MkMn0_transitions}
\end{theorem}

\begin{figure}[t!]
\centering
\medmuskip=0\medmuskip
\resizebox{1.04\textwidth}{!}{
\centering
~~~~\xymatrix{
*++[o][F]{0} \ar@/^2pc/[rr]^{\lambda} &
*++[o][F]{1} \ar@/^2pc/[rr]^{\lambda} \ar@/^1.5pc/[l]^{\mu}&
*++[o][F]{2} \ar@/^2pc/[rr]^{\lambda}
  \ar@/^1.5pc/[l]^{2\mu}&
*++[o][F]{3} \ar@/^2pc/[rr]^{\lambda} \ar@/^1.5pc/[l]^{3\mu}&
*++[o][F]{4} \ar@/^2pc/[rr]^{\lambda}
  \ar@/^1.5pc/[l]^{4\mu}&
*++[o][F]{5} \ar@/^2pc/[rr]^{\lambda} \ar@/^1.5pc/[l]^{4\mu}&
*++[o][F]{6} \ar@/^2pc/[rr]^{\lambda}
  \ar@/^1.5pc/[l]^{4\mu}&
*++[o][F]{7} \ar@/^2pc/[rr]^{\lambda} \ar@/^1.5pc/[l]^{4\mu}&
*++[o][F]{8} \ar@{halfFifty}@/^1.52pc/[rr]
  \ar@/^1.5pc/[l]^{4\mu}&
*++[o][F]{9} \ar@{halfForty}@/^1.52pc/[r] \ar@/^1.5pc/[l]^{4\mu}&
\cdots \ar@/^1.5pc/[l]^{4\mu} &
}
}
\caption{State transition diagram of the \MkMn(0) queue for $n=4$ and $k=2$. The notation at any state is the number of jobs $m$ in the system in that state. The set of boundary states are $\{0, 1, 2,3,4\}$, and the levels are pairs of states $\{5,6\}$, $\{7,8\}$, etc. The transition matrix is of the form~\eqref{eq:QBD_matrix} with $B_0 = [0~~0~~0~~0~~4\mu~;0~~0~~0~~0~~0]$, $B_1=[-\lambda~~0~~\lambda~~0~~0;~~\mu~~-(\mu+\lambda)~~0~~\lambda~~0;~~0~~2\mu~~-(2\mu+\lambda)~~0~~\lambda;~~0~~0~~3\mu~~-(3\mu+\lambda)~~0;~~0~~0~~0~~4\mu~~-(4\mu+\lambda)]$, $B_2=[0~~0~;~~0~~0~;~~0~~0~;~~\lambda~~0~;~~0~~\lambda]$, $A_0 = [0~~4\mu~;~~0~~0]$, $A_1=[-(4\mu+\lambda)~~0~;~~4\mu~~-(4\mu+\lambda)]$, $A_2=[\lambda~~0~;~~0~~\lambda]$. }
\label{fig:MkMn0_n4k2_statetransitiondiag}
\end{figure}
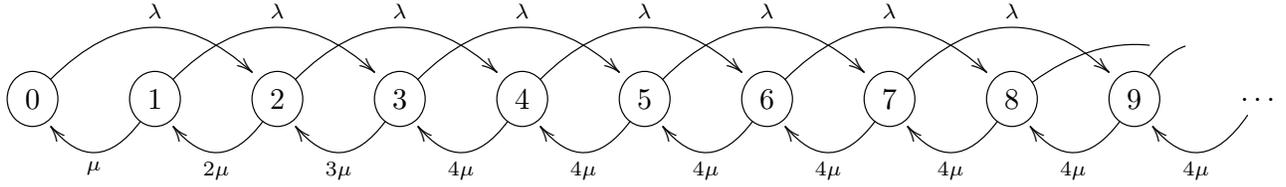

The state transition diagram of the \MkMn(0) queue for $n=4,k=2$ is shown in Fig.~\ref{fig:MkMn0_n4k2_statetransitiondiag}.

Theorem~\ref{thm:MDS_MkMn0_transitions} shows that the MDS-Reservation(0) queue is a QBD process, allowing us to employ the SMC solver to obtain its steady-state distribution. Alternatively, the \MkMn(0) queue is simple enough to allow for a direct analysis as well. Let $\pi_m$ denote the stationary probability of any state $m \in \{0,1,\ldots,\infty\}$. Then, for any $m \in \{1,\ldots,\infty\}$, the global balance equation for the cut between states $(m-1)$ and $m$ gives:
\beq \pi_m = \frac{\lambda}{\min(m, n)\mu} \sum_{j=(m-k)^+}^{m-1} \pi_j~.
\eeq
The stationary distribution of the Markov chain can now be computed easily from these equations.

\subsection{\MkMn(t) for a general t}
\subsubsection{Scheduling policy} 
Algorithm~\ref{alg:MkMnt} formally describes the \MkMn(t) scheduling policy.
\begin{algorithm}[h]
\caption{\MkMn(t) Scheduling Policy}
\begin{algorithmic}
\algostate \textbf{On} arrival of a batch
\algostate \algoindent \textbf{If}  {buffer has strictly fewer than t batches}
\algostate \algoindent \algoindent  Assign jobs of new batch to idle servers
\algostate \algoindent \textbf{Else if}  {buffer has t batches}
\algostate \algoindent \algoindent  Assign jobs of first batch to idle servers
\algostate \algoindent \algoindent \textbf{If} {first batch is cleared}
\algostate \algoindent \algoindent \algoindent Assign jobs of new batch to idle servers
\algostate \algoindent  Append remaining jobs of new batch to end of buffer
\algostate \textbf{On} departure of job from a server (say, server $s$)
\algostate \algoindent \textbf{If}  {number of batches in buffer is strictly greater than t}
\algostate \algoindent \algoindent  Assign job from first batch in buffer to this server
\algostate \algoindent \textbf{Else} 
\algostate \algoindent \algoindent  Among all batches in buffer that $s$ has not served, find the one that arrived earliest; assign a job of this batch to $s$
\end{algorithmic}
\label{alg:MkMnt}
\end{algorithm}

\begin{example} (t=1).
\begin{figure}[tb]
\centering
\subfloat[]{
\includegraphics[width=.16\textwidth]{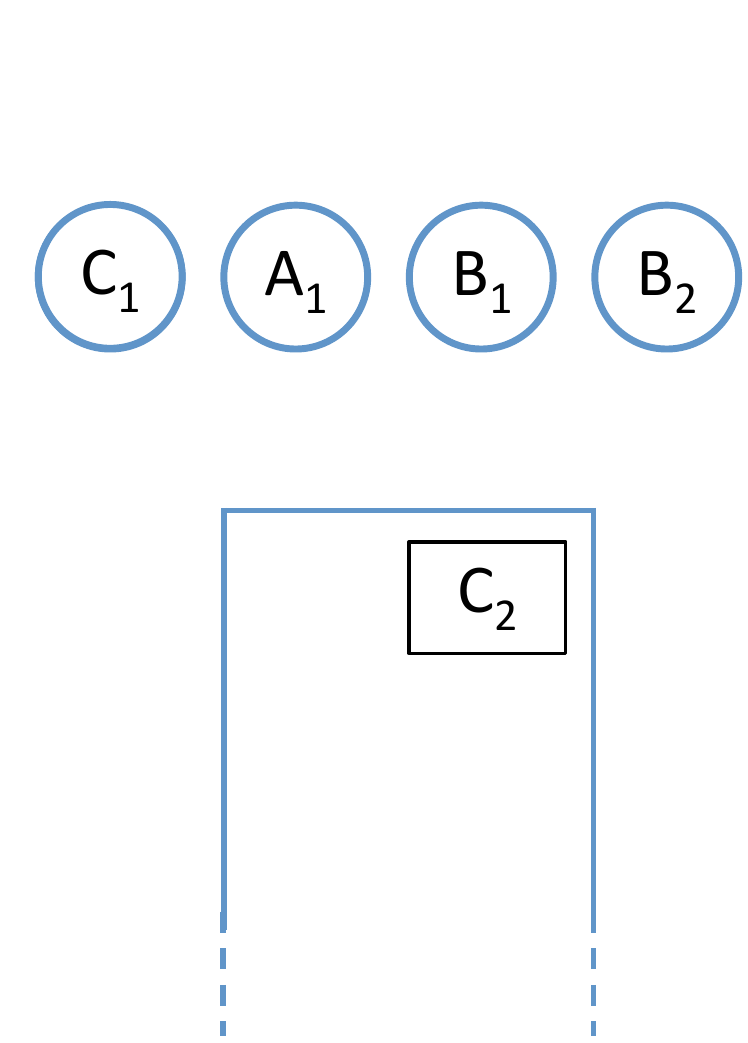}
\label{fig:MkMn1_workinga}
}~
\subfloat[]{
\includegraphics[width=.16\textwidth]{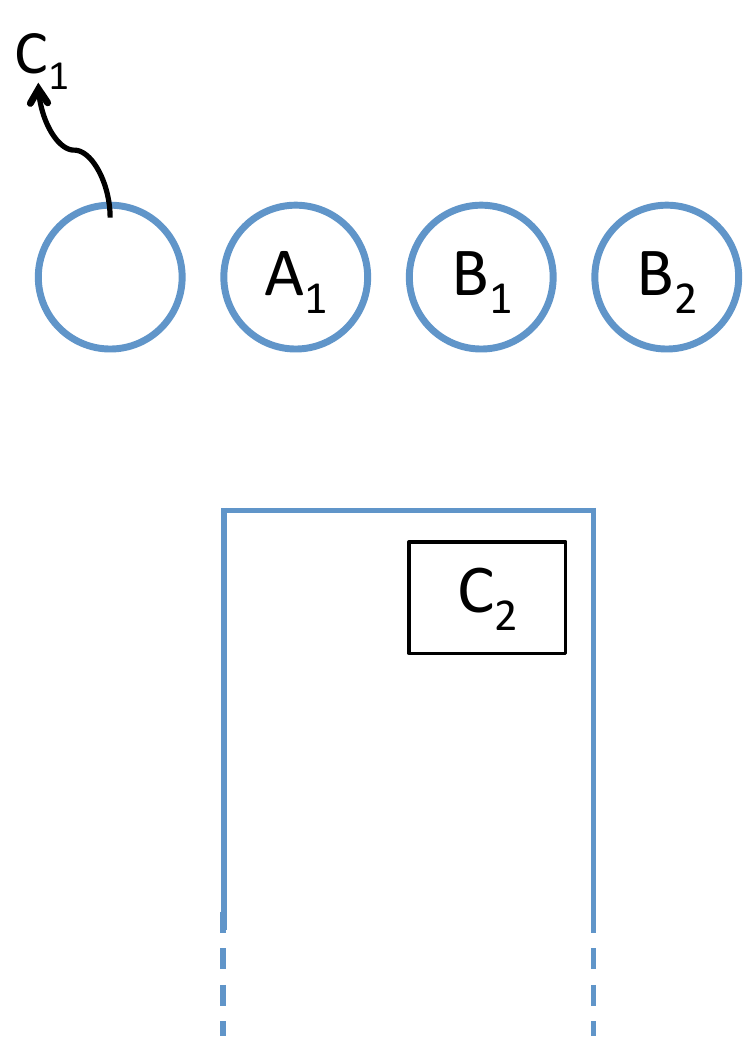}
\label{fig:MkMn1_workingb}
}~
\subfloat[]{
\includegraphics[trim=0 0 0 2cm, clip=true, width=.16\textwidth]{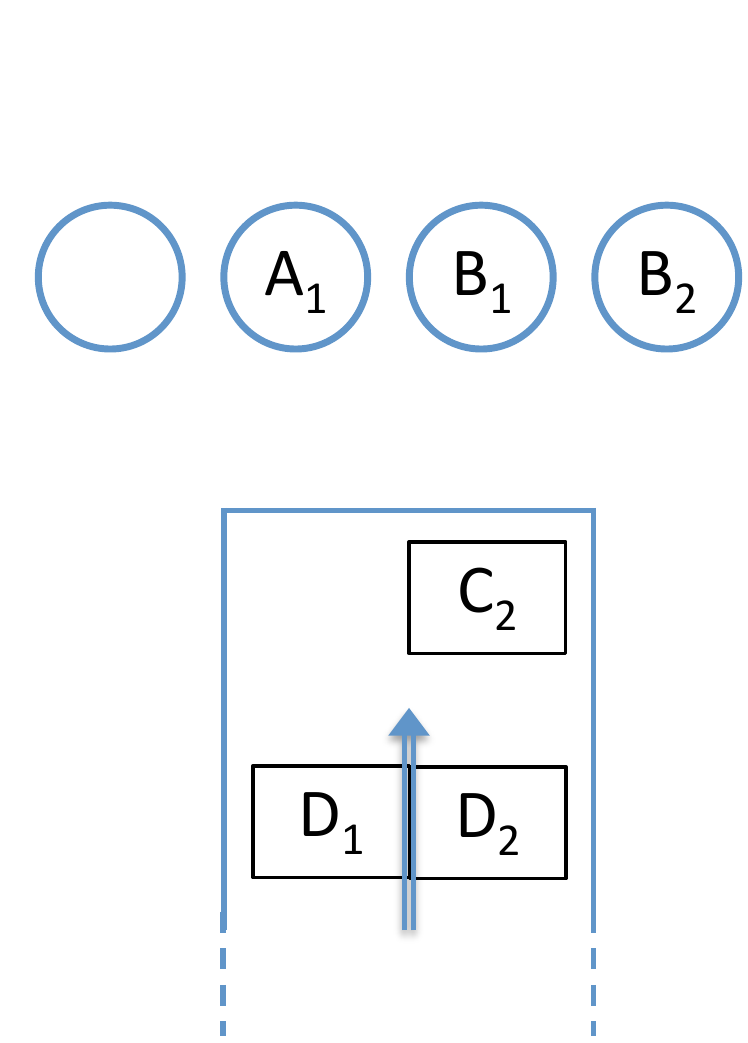}
\label{fig:MkMn1_workingc}
}~
\subfloat[]{
\includegraphics[trim=0 0 0 2cm, clip=true,width=.16\textwidth]{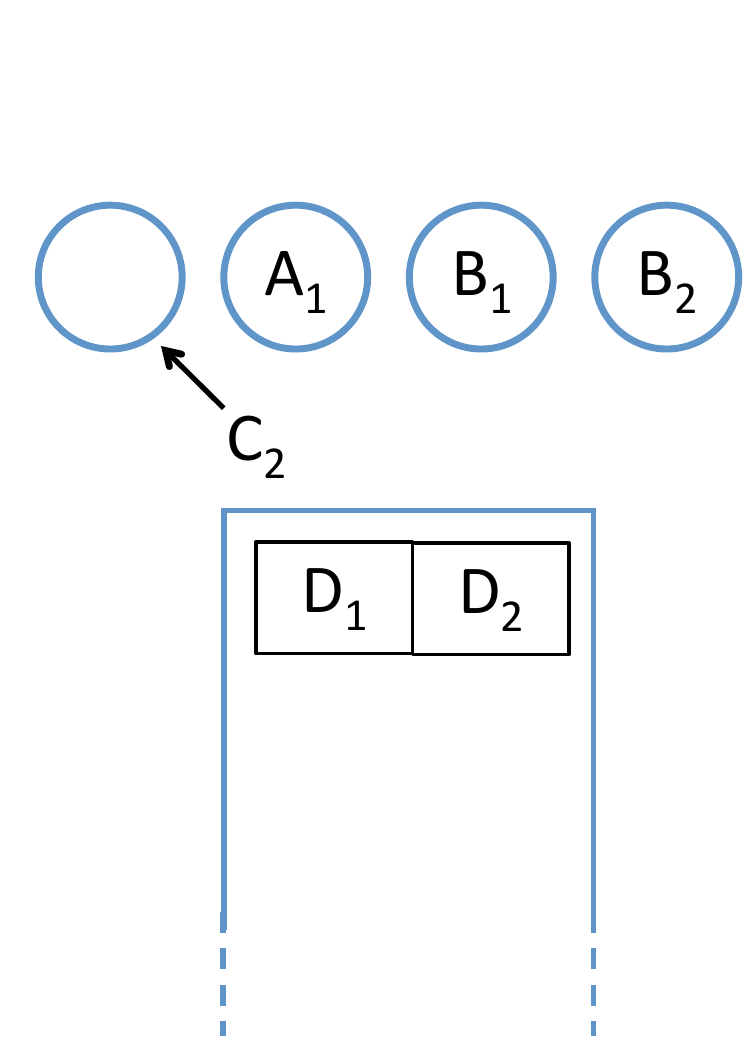}
\label{fig:MkMn1_workingd}
}
\caption{Illustration of the working of the \MkMn(1) scheduling policy. This policy allows a server to begin processing a job of a batch that it has already served, unless this batch is the only batch waiting in the buffer. As shown in the figure, server $1$ cannot process $C_2$ in (b) since it has already processed $C_1$ and $C$ is the only waiting batch; this restriction is removed upon on arrival of another batch in the buffer in (d).}
\label{fig:MkMn1_working}
\end{figure}
Consider a system in the state shown in Fig.~\ref{fig:MkMn1_workinga}. Suppose server $1$ completes execution of job $C_1$ (Fig.~\ref{fig:MkMn1_workingb}). In this situation, the processing of $C_2$ by server $1$ would be allowed under \MkMn(0), but prohibited in the MDS queue. The \MkMn(1) queue follows the scheduling policy of the MDS queue whenever the total number of batches in the buffer is no more than $1$, and hence in this case, server $1$ remains idle. Next, suppose there is an arrival of a new batch (Fig.~\ref{fig:MkMn1_workingc}). At this point there are two batches in the buffer, and the \MkMn(1) scheduling policy switches its mode of operation to allowing any server to serve any job. Thus, the first server now begins service of $C_2$ (Fig.~\ref{fig:MkMn1_workingd}).
\end{example}

\subsubsection{Analysis}
\begin{theorem}
The state transition diagram of the \MkMn(t) queue has a state space $\{0,1,\ldots,k\}^t \times \{0,1,2,\ldots\}$. It is a QBD process with boundary states $\{0,\ldots,k\}^t \times \{0,\ldots,n+tk\}$, and levels $\{0,\ldots,k\}^t \times \{n-k+1+jk,\ldots,n+jk\}$ for $j=\{t+1,t+2,\ldots,\infty\}$. 
\label{thm:MkMnt_transitions}
\end{theorem}
The proof of Theorem~\ref{thm:MkMnt_transitions} also describes how one can obtain the configuration of the entire system from only the number of jobs in the system, under the \MkMn(t) scheduling policies.

As in the case of MDS-Reservation(t) queues, one can see that the sequence of \MkMn(t) queues, as t increases, becomes closer to the MDS queue. On increase in the value of parameter t, the bounds become tighter, but the complexity of the transition diagrams also increases, and the limit of this sequence is the MDS queue itself.
\begin{theorem}
The \MkMn(t) queue, when t\,$=\infty$, is precisely the MDS(n,k) queue.
\label{thm:MkMn_infinity}
\end{theorem}

\begin{remark}The class of queues presented in this section have another interesting intellectual connection with the MDS queue:
the performance of an MDS(n,k) queue is lower bounded by the M\upseck/M/(n-k+1)(t) queue for all values of t.
\end{remark}

\section{Performance Comparison of Various Scheduling Policies}\label{sec:analysis}
In this section we analyse the performance of the MDS-Reservation(t) and the \MkMn(t) queues. The analysis is performed by first casting these queues as quasi-birth-death processes (as shown in Theorems~\ref{thm:MDS_Reservationt_transitions} and~\ref{thm:MkMnt_transitions}), and then building on the properties of quasi-birth-death processes to compute the average latency and throughput of each of these queues. Via simulations, we then validate these analytical results and also look at the performance of these queues with respect to additional metrics. We see that under each of these metrics, for the parameters simulated, the performance of the lower bound MDS-Reservation(t) queues (with $t=3$) closely follows the performance of the upper bound \MkMn(t) queues (with $t=1$).


\subsection{Maximum throughput}
The maximum throughput is the maximum possible number of requests that can be served by the system per unit time.
\begin{theorem}
\thickmuskip=.4\thickmuskip
\medmuskip=.4\medmuskip
Let $\rho^*_{\textrm{Resv(t)}}$, $\rho^*_{MDS}$, and $\rho^*_{M^k\!/M/n(t)}$ denote the maximum throughputs of the MDS-Reservation(t), MDS, and \MkMn(t) queues respectively. Then, 
\[\rho^*_{MDS} = \rho^*_{M^k\!/M/n(t)} = \frac{n}{k}~.\]
When $k$ is treated as a constant,
\[ \left(1-O(n^{-2})\right) \frac{n}{k} \leq \rho^*_{\textrm{Resv(t)}} \leq \frac{n}{k}~. \]
In particular, for any $t \geq 1$, when $k=2$,
\[ \left(1-\frac{1}{2n^2 - 2n +1}\right) \frac{n}{k} = \rho^*_{\textrm{Resv(1)}} \leq  \rho^*_{\textrm{Resv(t)}}~, \label{eq:throughput_k2}\]
and when $k=3$,
\[ \left(\!1-\frac{4n^3-8n^2+2n+4}{3n^5-12n^4+22n^3-29n^2+26n-8}\!\right) \frac{n}{k} = \rho^*_{\textrm{Resv(1)}} \leq  \rho^*_{\textrm{Resv(t)}} \label{eq:throughput_k3} \]
\label{thm:throughput}
\end{theorem}

Note that the special case of Theorem~\ref{thm:throughput}, for the MDS-Reservation(1) queue with $k=2$, also recovers the throughput results of~\cite{huang2012codes}. Moreover, as compared to the proof techniques of~\cite{huang2012codes}, the proofs in this paper are much simpler and do not require computation of the stationary distribution. Using the techniques presented in the proof of Theorem~\ref{thm:throughput}, explicit bounds analogous to those for $k=2,\ 3$ in Theorem~\ref{thm:throughput} can be computed for $k\geq 4$ as well.

\begin{figure}[t!]
\begin{minipage}{.47\textwidth}
\centering
\includegraphics[width=.95\textwidth]{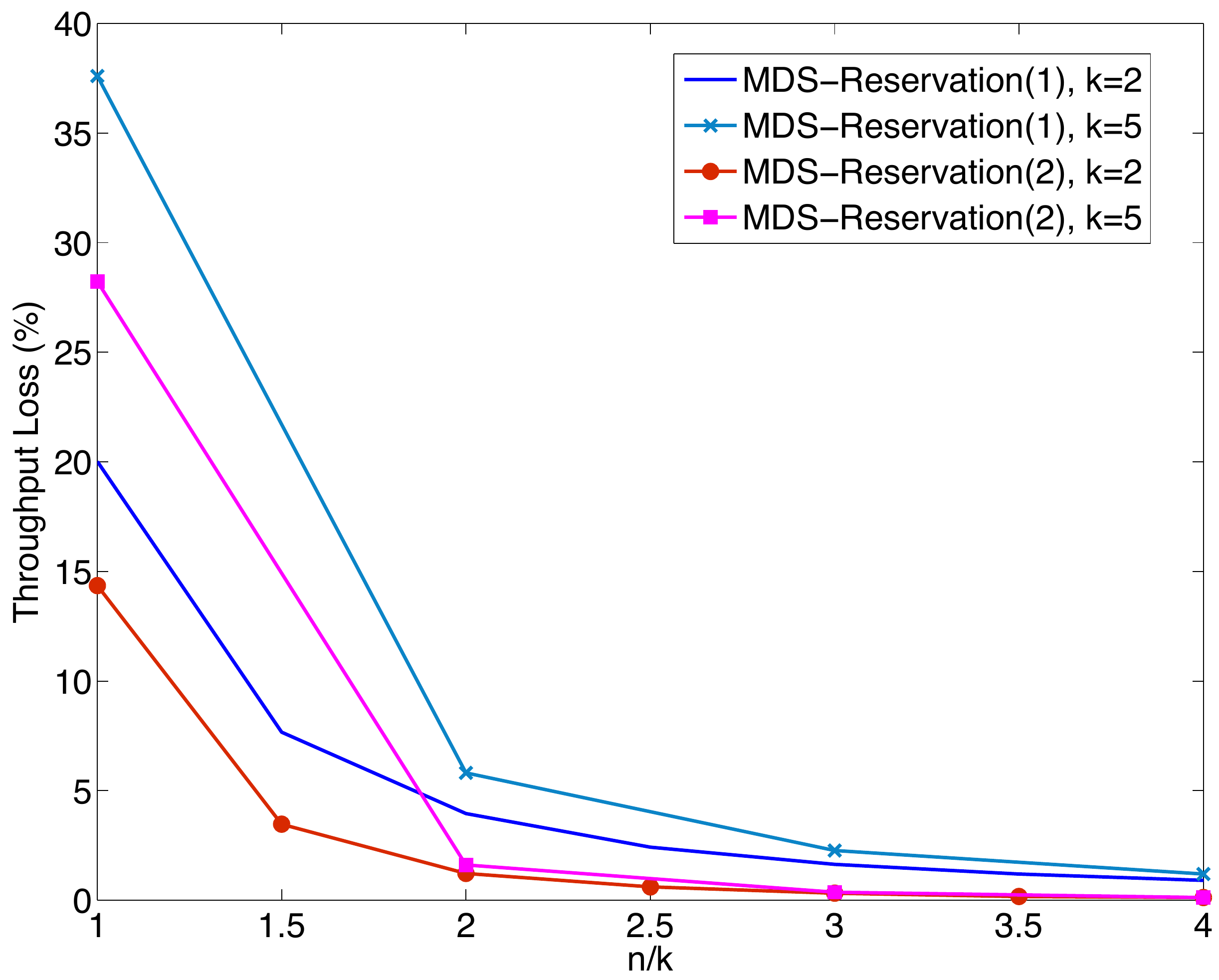}
\caption{Loss in maximum throughput incurred by the MDS-Reservation(1) and the MDS-Reservation(2) queues as compared to that of the MDS queue.}
\label{fig:plot_throughput_loss}
\end{minipage}
\hspace{.05\textwidth}
\begin{minipage}{.47\textwidth}
\centering
\includegraphics[width=\textwidth]{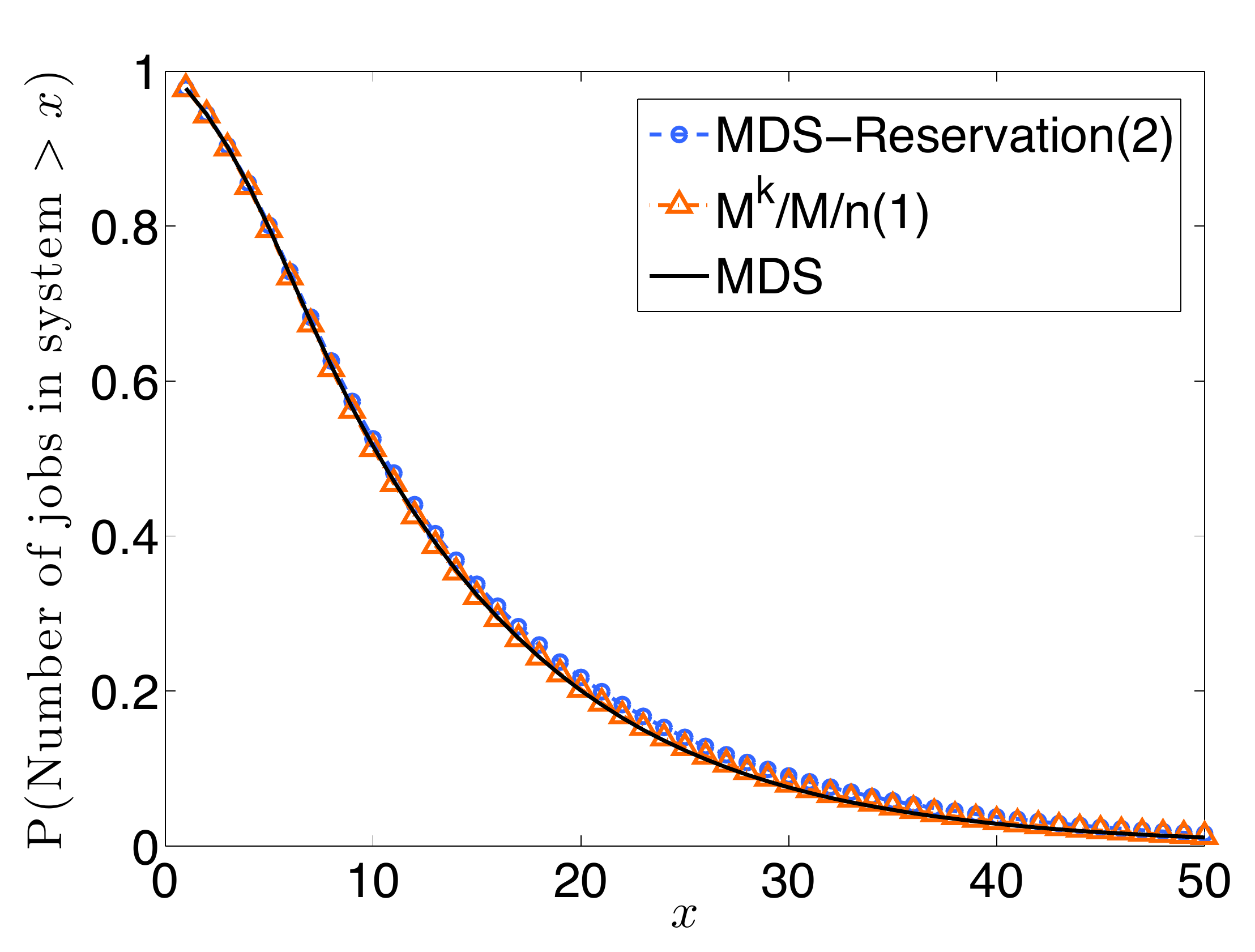}
\caption{Steady state distribution (complementary cdf) of the system occupancy when $n=10$, $k=5$, $\lambda=1.5$ and $\mu=1$.}
\label{fig:plot_system_occupancy}
\end{minipage}\\~\\~\\
\begin{minipage}{.47\textwidth}
\centering
\includegraphics[width=.95\textwidth]{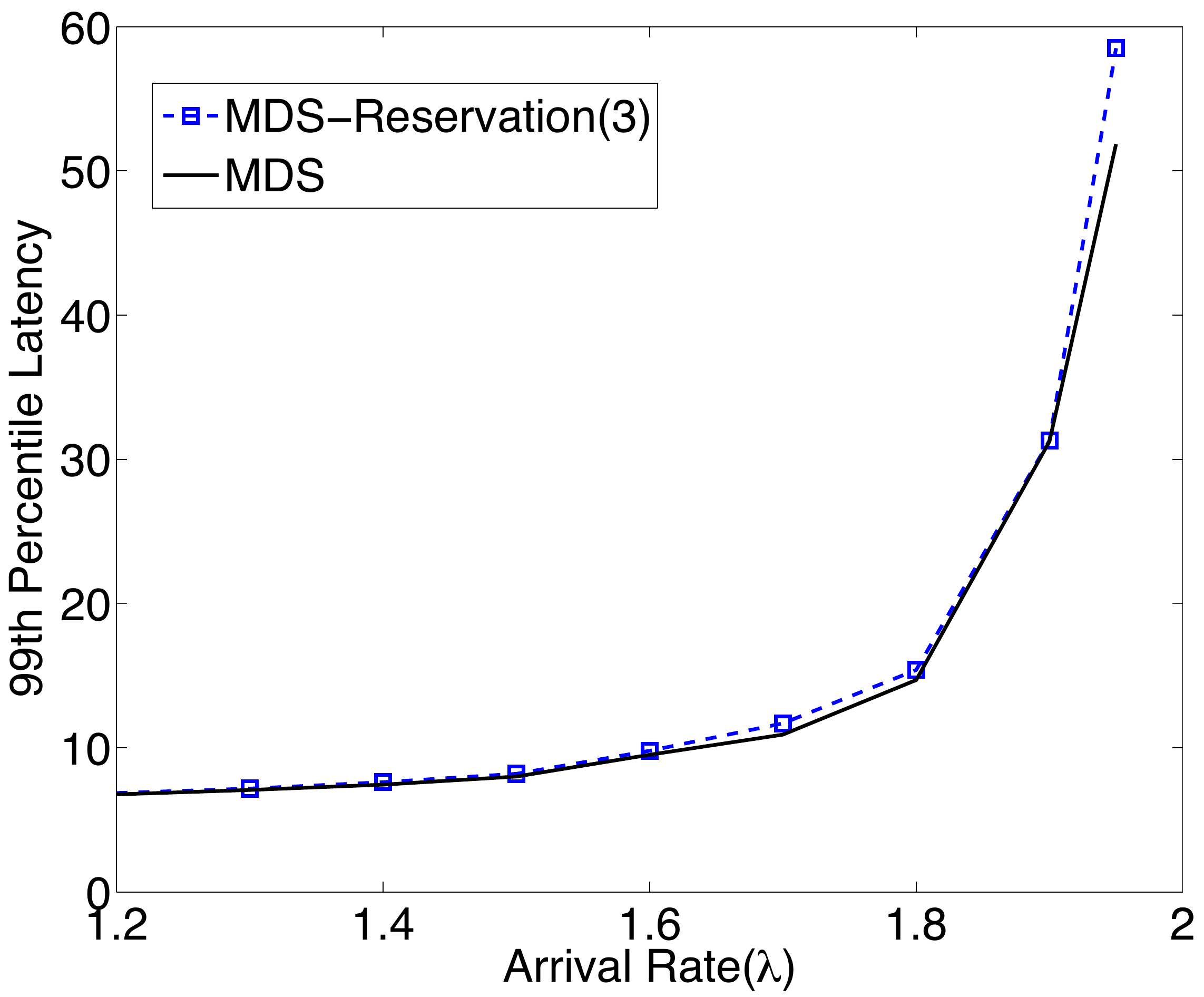}
\caption{The $99\supth$ percentile of the distribution of the latency in a system with parameters $n=10$, $k=5$ and $\mu=1$. For any arrival rate $\lambda$, a curve takes value $y(\lambda)$ if exactly $1\%$ of the batches incur a delay greater than $y(\lambda)$}
\label{fig:plot_tail_batch_latency}
\end{minipage}\hspace{.05\textwidth}
\begin{minipage}{.47\textwidth}
\centering
\includegraphics[width=\textwidth]{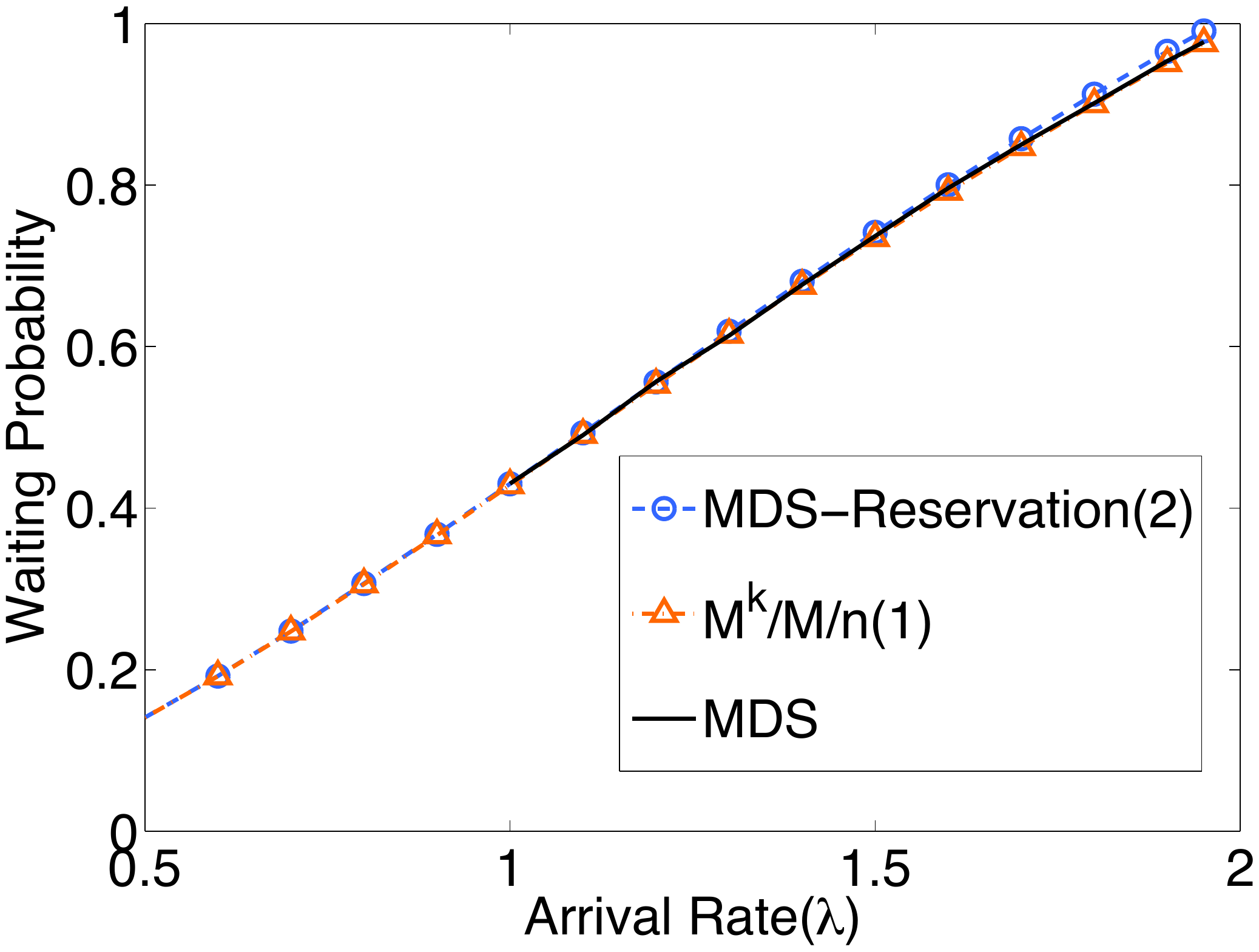}
\caption{Waiting probability in a system with $n=10$, $k=5$ and $\mu=1$.}
\label{fig:plot_waiting_probability}
\end{minipage}

\end{figure}

Fig.~\ref{fig:plot_throughput_loss} plots loss in maximum throughput incurred by the MDS-Reservation(1) and the MDS-Reservation(2) queues, as compared to that of the MDS queue.

\subsection{System occupancy}
The system occupancy at any given time is the number of jobs present (i.e., that have not yet been served completely) in the system at that time. This includes jobs that are waiting in the buffer as well as the jobs being processed in the servers at that time. The distribution of the system occupancy is obtained directly from the stationary distribution of the Markov chains constructed in sections~\ref{sec:lower} and~\ref{sec:upper}. Fig.~\ref{fig:plot_system_occupancy} plots the complementary cdf of the the number of jobs in the system in the steady state. Observe that the analytical upper and lower bounds from the MDS-Reservation(2) and the \MkMn(1) queues respectively are very close to each other. 

\subsection{Latency}
The latency faced by a batch is the time from its arrival into the system till the completion of service of $k$ of its jobs. 

{\it Average:}
We have analytically computed the average batch latencies of the MDS-Reservation(t) and the \MkMn(t) queues in the following manner. We first compute the steady state distribution $\mathbf{\pi}$ of the number of jobs in the system. As discussed in the previous sections, each state $i$ in the corresponding Markov chain is associated to a unique configuration of the jobs in the system. Now, the average latency $d_i$ faced by a batch entering when the system is in state $i$ can be computed easily by employing the Markovian representations of the queues presented previously. Since Poisson arrivals see time averages~\cite{wolff1982poisson}, the average latency faced by a batch in the steady state is given by $\sum_{i} \pi_i d_i$.

Fig.~\ref{fig:preview} plots the average latency faced by a batch in the steady state. Observe that the performance of the MDS-Reservation(t) scheduling policy, for t as small as $3$, is extremely close to that of the MDS scheduling policy and to the upper bounding \MkMn(1) scheduling policy.

{\it Tails (simulations only):}
We also look at the tails of the latency-distribution via simulations. Fig.~\ref{fig:plot_tail_batch_latency} plots the $99\supth$ percentile of the distribution of the latency. Also observe how closely the bound MDS-Reservation(3) follows the exact MDS.

\subsection{Waiting probability}
The waiting probability is the probability that, in the steady state, one or more jobs of an arriving batch will have to wait in the buffer and will not be served immediately. As shown previously, under the MDS-Reservation(t) and the \MkMn(t) scheduling policies, the system configuration at any time is determined by the state of the Markov chain, which also determines whether a newly arriving batch needs to wait or not. Thus, the waiting probability can be computed directly from the steady state distribution of the number of jobs in the system. Fig.~\ref{fig:plot_waiting_probability} plots the waiting probability for the different queues considered in the paper. Observe how tightly the \MkMn(1) and the MDS-Reservation(2) queues bound the waiting probability of the MDS queue.

\subsection{Degraded Reads}
The system considered so far assumed that each incoming request desires one entire file (from any $k$ servers). In certain applications, incoming requests may sometimes require only a part of the file, say, the part that is stored in one of the servers. In the event that a server is busy or has failed, one may serve requests for that part from the data stored in the remaining servers. This operation is termed a `degraded read'. Under an MDS code, a degraded read may be performed by obtaining the entire file from the data stored in any $k$ of the $(n-1)$ remaining servers, and then extracting the desired part. Such an operation is generally called `data-reconstruction'.

Dimakis et al. recently proposed a new model, called the \textit{regenerating codes model}, as a basis to design alternative codes supporting faster degraded reads. Several explicit codes under this model have been constructed subsequently, e.g.,~\cite{rashmi2009allerton,rashmi2011productmatrix,papailiopoulos2011repair,tamo2012access}. In particular, the \textit{product-matrix (PM) codes} proposed in~\cite{rashmi2011productmatrix} are practical codes that possess several appealing properties. One feature of the product-matrix codes is that they are associated to an additional parameter $d~(\leq n-1)$, and can recover the desired data by reading and downloading a fraction ${\frac{1}{d-k+1}}^{\textrm{th}}$ of the requisite data from \textbf{any} $d$ of the remaining $(n-1)$ servers. This method of performing degraded reads entails a smaller total download but requires connectivity to more nodes, and hence the gains achieved by this method in a dynamic setting are unclear. This operation is generally called a `node-repair' operation in the literature.

\begin{figure}[t]
\centering
\includegraphics[width=.6\textwidth]{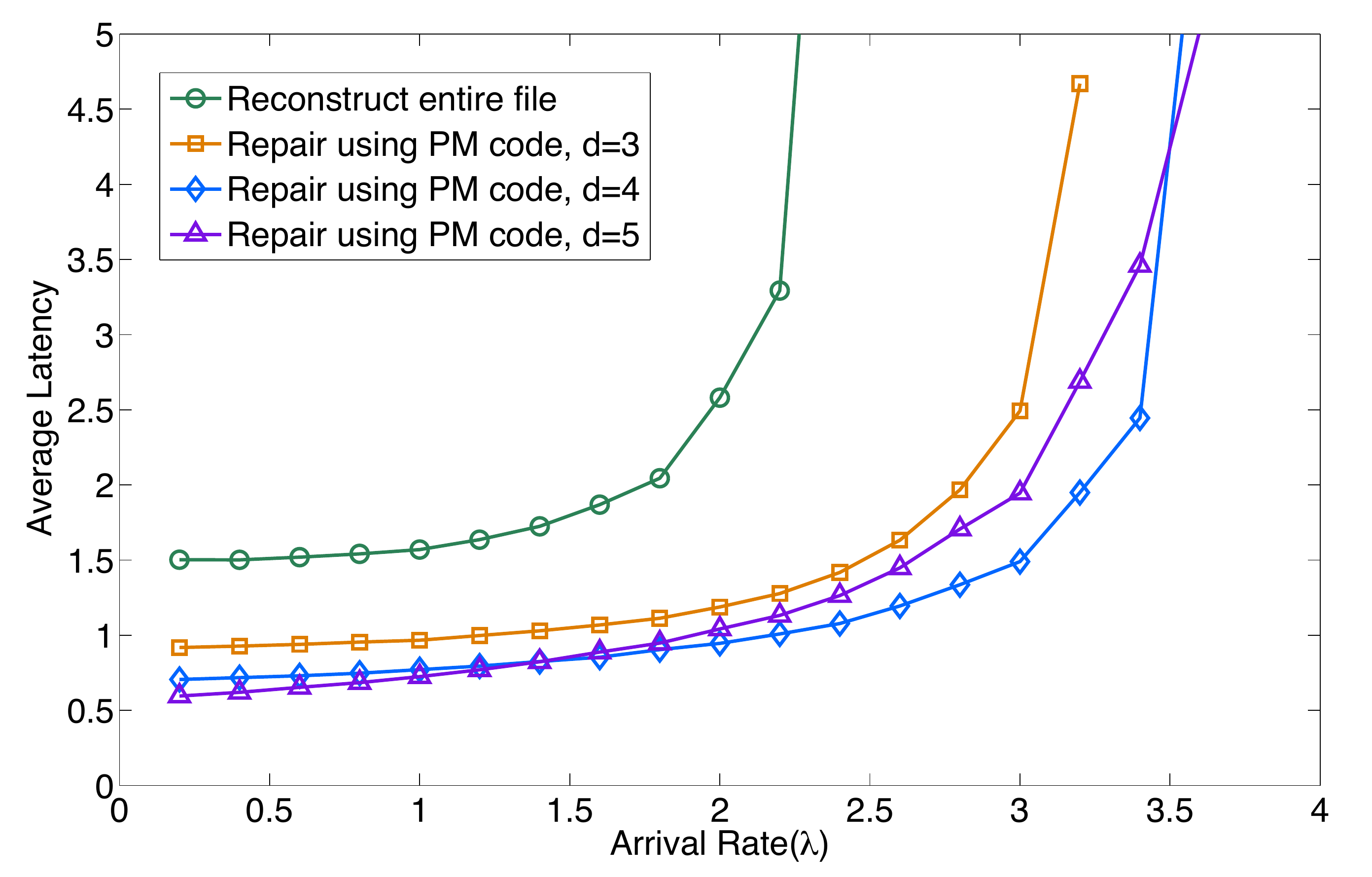}
\caption{Average latency during degraded reads. The parameters associated to this system are $n=6$, $k=2$. The service time at any server is exponentially distributed with a mean proportional to the amount of data to be read. The performance of the product-matrix (PM) codes are compared for various values of the associated parameter $d$.}
\label{fig:plot_partial}
\end{figure}

We employ the framework of the MDS queue to compare these two methods of performing degraded reads: the data-reconstruction method results in an MDS(n-1,k) queue, while the node-repair operation leads to an MDS(n-1,d) queue (with a different service rate). Fig.~\ref{fig:plot_partial} depicts average latency incurred under the two methods when $n=6$, $k=2$ and $d=3$. We see that the product-matrix/regenerating codes perform consistently better in terms of the latency performance. The key insight is that the property of being able to read from \textbf{any} $d$ servers provides a great deal of flexibility to the degraded-read operations under a product-matrix code, enabling it to match up (and beat) the performance of the data-reconstruction operation.


\section{Discussion and Open Problems}\label{sec:future}
We aim to use the results of this paper as a starting point for analysis of more complex systems that mimic the real world more closely. In particular, we intend to build upon the ``MDS-Queue'' framework presented here to analyse queues that relax one or more of the assumptions made in the paper, e.g., having general service times, heterogeneous requests or servers, non-MDS codes, and decentralized queues. 

In this paper, while we characterized the metrics of average latency and throughput analytically, we also used simulations to examine the performance of the queues in terms of several additional metrics. We observed in these simulations that the MDS-Reservation(t) and the \MkMn(t) scheduling policies result in a performance very close to that of the exact MDS queue for values of t as small as $3$. Thus, even respect to these metrics, an analysis of the (simpler) scheduling policies of MDS-Reservation(t) and \MkMn(t) may provide rather-accurate estimations of the performance of MDS codes.

\bibliographystyle{IEEEtran}

\appendix[Proofs of Theorems]

\textbf{\textit{Additional Notation used in the proofs:}}
Table~\ref{table:notation} enumerates notation for various parameters that describe the system at any given time. To illustrate this notation, consider again the system depicted in Fig.~\ref{fig:MDSqueue_working_a}. Here, the parameters listed in Table~\ref{table:notation} take values $m=10$, $z=0$, $b=3$, $s_1=0$, $s_2=0$, $s_3=0$, $w_1=2$, $w_2=2$ and $w_3=2$. One can verify that keeping track of these parameters leads to a valid Markov chain (under each of the scheduling policies discussed in this paper). Note that we do not keep track of the jobs of a batch once all $k$ jobs of that batch have begun to be served, nor do we track what servers are serving what jobs. This is to ensure a smaller complexity of representation an2d computation. Further note that in terms of the parameters listed in Table~\ref{table:notation}, the number of servers that are busy at any given time is equal to $(n-z)$. For batch $i$ in the buffer ($i \in \{1,\ldots,b\}$), the number of jobs that have completed service is equal to $(k-s_i-w_i)$. For any integer $i$, $w_i=0$ will mean that there is no $i\supth$ waiting batch in the buffer.

\begin{table}[b]
\centering
\normalsize
\begin{tabular}{|c|l|l|}
\hline
Value & Meaning  & Range\\
\hline
$m$ & number of jobs in the entire system & $0$ to $\infty$\\
$z$& number of idle servers & $0$ to $n$ \\
$b$ & number of waiting batches & $0$ to $\infty$\\
$\{s_i\}_{i=1}^{b}$ & number of of jobs of $i^{\textrm{th}}$ waiting batch, in the servers  & $0$ to $k-1$\\
$\{w_i\}_{i=1}^{b}$ & number of of jobs of $i^{\textrm{th}}$ waiting batch, in the buffer &  $0$ to $k$\\
\hline
\end{tabular}
\caption{\normalsize Notation used to describe state of the system.}
\label{table:notation}
\end{table}

\begin{IEEEproof}[\textbf{Proof of Theorem~\ref{thm:MDS_Reservation0_transitions}}]
Since the scheduling policy mandates all $k$ jobs of any batch to start service together, the number of jobs in the buffer is necessarily a multiple of $k$. Furthermore, when the buffer is not empty, the number of servers that are idle must be strictly smaller than $k$ (since otherwise, the first waiting batch can be served). It follows that when $m\leq n$, the buffer is empty ($b=0$), and all $m$ jobs are being served by $m$ servers (and $z=(n-m)$ servers are idle). When $m > n$, the buffer is not empty. Assuming there are $z$ idle servers, there must be $(n-z)$ jobs currently being served, and hence there are $m-(n-z)$ jobs in the buffer. However, since the number of jobs in the buffer must be a multiple of $k$, and since $z \in \{0,1,\ldots,k-1\}$, it must be that $z = (n-m) \modulo k$. Thus, when $m>n$, there are $b=\frac{m-n+z}{k}$ batches waiting in the buffer, and $w_i=k,~s_i=0~\forall~i \in \{1,\ldots,b\}$. We have thus shown that the knowledge of $m$ suffices to completely describe the system.

Once we have determined the configuration of the system as above, it is now easy to obtain the transitions between the states. An arrival of a batch increases the total number of jobs in the system by $k$, and hence the transition from state $m$ to $(m+k)$ at rate $\lambda$. When $m \leq n$, all $m$ jobs are being served, and the buffer is empty. Thus, the total number of jobs in the system reduces to $(m-1)$ at rate $m\mu$. When $m > n$, the number of jobs being served is $n-z = n- ((n-m) \modulo k)$, and thus there is a transition from state $m$ to $(m-1)$ at rate $(n- ((n-m) \modulo k))\mu$.
\end{IEEEproof}

\begin{IEEEproof}[\textbf{Proof of Theorem~\ref{thm:MDS_Reservation1_transitions}}]
Results as a special case of Theorem~\ref{thm:MDS_Reservationt_transitions}. As a side note, in any given state $(w_1 , m) \in \{0,1,\ldots,k\} \times \{0,1,\ldots,\infty\}$ of the resulting Markov chain, the number of idle servers is given by $z=n-m$ if $m \leq n-k$, and $z=(n+w_1-m)\modulo k$ otherwise. The state $(w_1,m)$ has transitions to state:
\begin{itemize}
\itemsep0em
\item $( (m+k-n)^+ , m+k)$ at rate $\lambda$, if $w_1= 0$.
\item $(w_1 , m + k)$ at rate $\lambda$, if $w_1 \neq 0$
\item $(w_1 , m-1)$ at rate $m\mu$, if $w_1= 0$.
\item $(w_1 , m-1)$ at rate $(k-w_1-z)\mu$, if $w_1\neq 0$
\item $(w_1-1 , m-1)$ at rate $(n-k+w_1)\mu$, if $(w_1 > 1$  or $(w_1=1$ \& $m\leq n+1))$
\item $((k-z)^+ , m-1)$ at rate $(n-k+w_1)\mu$, if $(w_1 = 1$ \& $m > n+1)$.
\end{itemize}
\end{IEEEproof}

\begin{IEEEproof}[\textbf{Proof of Theorem~\ref{thm:MDS_Reservationt_transitions}}]
For any state of the system $(w_1,w_2,\ldots,w_t, m) \in \{0,1,\ldots,k\}^t \times \{0,1,\ldots,\infty\}$, define
\beq q = \begin{cases} 
0 & \textrm{if~~} w_1 = 0\\
t & \textrm{else if~~} w_t \neq 0\\
\text{arg max} \{t': w_{t'} \neq 0,\, 1\leq t' \leq t\} ~~~& \textrm{otherwise}.\\
\end{cases}\label{eq:MDS_Reservationt_transitions_q}\eeq
It can be shown that
\beq b = \begin{cases}
0 & \textrm{if~~} q=0\\
q & \textrm{if~~} 0<q<t\\
t+\left\lfloor \frac{m- \sum_{j=1}^{t}w_i-n}{k}\right\rfloor ~~& \textrm{otherwise},
\end{cases}
\eeq
\beq z=n-(m- \sum_{j=1}^{t}w_j-(b-t)^+k)~,\eeq
\beq \textrm{for~} i \in \{1,\ldots,b\},~~
 s_i = \begin{cases}
w_{i+1}-w_i & \textrm{if~~} i \in \{1,\ldots,q-1\}\\
k-z-w_q & \textrm{if~~} i=q \\
0 &\textrm{if~~} i \in \{q+1,\ldots,b\},
\end{cases}
\eeq
and \beq \textrm{for~} i \in \{t+1,\ldots,b\},~~w_i=k~.\eeq
Given the complete description of the state of the system as above, the characterization of the transition diagram is a straightforward task. 

It is not difficult to see that the MDS-Reservation(t) queue has the following two key features: (a) any transition event changes the value of $m$ by at most $k$, and (b) for $m \geq n-k+1+tk$, the transition from any state $(w_1,\,m)$ to any other state $(w_1',\,m'\geq n-k+1+tk)$ depends on $m\modulo k$ and not on the actual value of $m$. This results in a QBD process with boundary states and levels as specified in the statement of the theorem. Intuitively, this says that when $m\geq n-k+1+tk$, the presence of an additional batch at the end of the buffer has no effect on the functioning of the system. (In contrast, when $m<n-k+1+tk$, the system may behave differently if there was to be an additional batch, due to the possibility of this batch being within the threshold t. For instance, a job of this additional batch may be served upon completion of service at a server, which is not possible if this additional batch was not present).

\end{IEEEproof}

\begin{IEEEproof}[\textbf{Proof of Theorem~\ref{thm:MDS_Reservation_infinity}}]
The MDS-Reservation(t) scheduling policy treats the first t waiting batches in the buffer as per the MDS scheduling policy, while imposing an additional restriction on batches $(t+1)$ onwards. When t$=\infty$, every batch is treated as in MDS, thus making MDS-Reservation($\infty$) identical to MDS.
\end{IEEEproof}

\begin{IEEEproof}[\textbf{Proof of Theorem~\ref{thm:MDS_MkMn0_transitions}}]
Under \MkMn(0), any job can be processed by any server, and hence a server may be idle only when the buffer is empty. Thus, when $m\leq n$, all $m$ jobs are in the servers, and the buffer is empty. When $m>n$, all the $n$ servers are full and the remaining $(m-n)$ jobs are in the buffer. The transitions follow as a direct consequence of these observations. It also follows that when $m\leq n$, $b=0$ and $z=n-m$.  In addition, in state $m\,(>n)$ it must be that $w_1 = (m-n)\textrm{mod}k$, $b=\lceil\frac{m-n}{k}\rceil$, and for $i\in\{2,\ldots,b\}$, $w_i=k$. Thus the knowledge of $m$ suffices to describe the configuration of the entire system.
\end{IEEEproof}

\begin{IEEEproof}[\textbf{Proof of Theorem~\ref{thm:MkMnt_transitions}}]
\thinmuskip=0\thinmuskip
For any state $(w_1,w_2,\ldots,w_t, m)$, define $q$ as in~\eqref{eq:MDS_Reservationt_transitions_q}. The values of $b$, $z$, $w_i$, are identical to that in the proof of Theorem~\ref{thm:MDS_Reservationt_transitions}. Given the complete description of the state of the system as above, the characterization of the transition diagram is a straightforward task. It is not difficult to see that the \MkMn(t) queues have the following two key features: (a) any transition event changes the value of $m$ by at most $k$, and (b) for $m \geq n+1+tk$, the transition from any state $(w_1,\,m)$ to any other state $(w_1',\,m'\geq n+1+tk)$ depends on $m\modulo k$ and not on the actual value of $m$. This results in a QBD process with boundary states and levels as specified in the statement of the theorem. Intuitively, this says that when $m\geq n+1+tk$, the total number of waiting batches is strictly greater than t. In this situation, the presence of an additional batch at the end of the buffer has no effect on the functioning of the system.
\end{IEEEproof}

\begin{IEEEproof}[\textbf{Proof of Theorem~\ref{thm:MkMn_infinity}}]
The \MkMn(t) scheduling policy follows the MDS scheduling policy when the number of batches in the buffer is less than or equal to t. Thus, \MkMn($\infty$) is always identical to MDS.
\end{IEEEproof}

\begin{IEEEproof}[\textbf{Proof of Theorem~\ref{thm:throughput}}]
In the MDS queue, suppose there are a large number of batches waiting in the buffer. Then, whenever a server completes a service, one can always find a waiting batch that has not been served by that server. Thus, no server is ever idle. Since the system has $n$ servers, each serving jobs with times i.i.d. exponential with rates $\mu$, the average number of jobs exiting the system per unit time is $n\mu$. The above argument also implies that under no circumstances (under any scheduling policy), can the average number of jobs exiting the system per unit time exceed $n\mu$. Finally, since each batch consists of $k$ jobs, the rate at which batches exit the system is $\lambda^*_{MDS}=\frac{n\mu}{k}$ per unit time. Since the \MkMn(t) queues upper bound the performance of the MDS queue, $\lambda^*_{\textrm{M$^k$/M/n(t)}}=\frac{n\mu}{k}$ for every $t$.
 
We shall now evaluate the maximum throughput of MDS-Reservation(1) by exploiting properties of QBD systems. In general, the maximum throughput $\lambda^*$ of any QBD system is the value of $\lambda$ such that: $\exists~\textbf{v}$ satisfying $\mathbf{v}^T(A_0 + A_1 + A_2) = 0$ and $\mathbf{v}^TA_0\mathbf{1} = \mathbf{v}^TA_2\textbf{1}$, where $\mathbf{1}=[1~1~\cdots~1]^T$. Note that the matrices $A_0$, $A_1$ and $A_2$ are affine transformations of $\lambda$ (for fixed values of $\mu$ and $k$). Using the values of $A_0$, $A_1$, $A_2$ in the QBD representation of MDS-Reservation(1), we can show that $\lambda^*_{\textrm{Resv(1)}} \geq (1-O(n^{-2}))\frac{n}{k}\mu$. For t$\geq2$, each of the MDS-Reservation(t) queues upper bound MDS-Reservation(1), and are themselves upper bounded by the MDS queue. It follows that  $\frac{n}{k}\mu \geq \lambda^*_{\textrm{Resv(t)}} \geq (1-O(n^{-2}))\frac{n}{k}\mu$ for t$ \geq 1$.  

The value of $\lambda^*_{\textrm{Resv(t)}}$ can be explicitly computed for any value of $n$, $k$ and t via the method described above. We perform this computation for $k=2$ and $k=3$ when t$=1$ to obtain the result mentioned in the statement of the theorem. We show the computation for $k=2$ here.

When $k=2$ and t$=1$, the $j\supth$ level of the QBD process consists of states $\{0,1,2\}\times\{n-1+2j,n+2j\}$ for $j\geq 1$. However, as seen in Fig.~\ref{fig:MDS_Reservation1_n4k2_statetransitiondiag}, several of these states never occur. In particular, in level $j$, only the states $(1,n-1+2j)$, $(1,n+2j)$ and $(2,n+2j)$ may be visited. Thus, to simplify notation, in the following discussion we consider the QBD process assuming the existence of only these three states (in that order) in every level. Under this representation, we have
\[
A_0 = \begin{bmatrix}
	0 & \mu & (n-1)\mu \\[0.3em]
	0 & 0 & 0 \\[0.3em]
	0 & 0 & 0
	\end{bmatrix}
, ~~A_1 = \begin{bmatrix}
	-n\mu-\lambda & 0 & 0 \\[0.3em]
	(n-1)\mu & -(n-1)\mu-\lambda & 0 \\[0.3em]
	n\mu & 0 & -n\mu-\lambda
	\end{bmatrix}
, ~~A_2 = \begin{bmatrix}
	\lambda & 0 & 0 \\[0.3em]
	0 & \lambda & 0 \\[0.3em]
	0 & 0 & \lambda \\[0.3em]
	\end{bmatrix}.
\]
\beq
\Rightarrow A_0 + A_1 + A_2 = \begin{bmatrix}
	-n & 1 & n-1\\[0.3em]
	n-1 & -(n-1) & 0\\[0.3em]
	n & 0 & -n
		\end{bmatrix} \mu
\eeq
One can verify that the vector
\[ \mathbf{v} = \begin{bmatrix}
	v_1 & v_2 & v_3
	\end{bmatrix}^T
	= \begin{bmatrix}
	n-1 ~&~ 1~&~\frac{(n-1)^2}{n} 
	\end{bmatrix}^T
	\]
satisfies \beq \mathbf{v}^T(A_0 + A_1 + A_2) = 0.\eeq Thus,
\beq \mathbf{v}^T A_0\mathbf{1} = n(n-1)\mu,\eeq and \beq \mathbf{v}^TA_2\mathbf{1} = \lambda \left(n + \frac{(n-1)^2}{n}\right)~.\eeq
According to the properties of QBD processes, the value of $\lambda = \lambda^*_{\textrm{Resv(1)}} $ must satisfy $\mathbf{v}^T A_0\mathbf{1} =\mathbf{v}^T A_2\mathbf{1}$. Thus,
\beq \lambda^*_{\textrm{Resv(1)}} = \frac{n^2(n-1)}{2n^2 - 2n + 1} = \left(1-\frac{1}{2n^2 - 2n + 1}\right)\frac{n}{2}\mu.
\eeq
\comment{
Proof for $k=3$:
\begin{align*}
A_0 &= \begin{bmatrix}
	0 & 0 & (n-2)\mu & 0 & 0 & 0 \\[0.3em]
	0 & 0 & 0 & 0 & 0 & 0 \\[0.3em]
	0 & 0 & 0 & 0 & 0 & 0 \\[0.3em]
	0 & (n-2)\mu & 0 & 0 & 0 & 0 \\[0.3em]
	0 & 0 & 0 & 0 & 0 & 0 \\[0.3em]
	(n-2)\mu & 0 & 0 & 0 & 0 & 0 \\[0.3em]
	\end{bmatrix},\\
A_1 &= \begin{bmatrix}
	-n\mu -\lambda & 0 & 0 & 2\mu & 0 & 0\\[0.3em]
	(n-1)\mu & -n\mu-\lambda & 0 & 0 & \mu & 0\\[0.3em]
	0 & n\mu & -n\mu-\lambda & 0 & 0 & 0\\[0.3em]
	0 & 0 & 0 & -(n-1)\mu-\lambda & 0 & \mu\\[0.3em]
	0 & 0 & 0 & (n-1)\mu & -(n-1)\mu-\lambda & 0\\[0.3em]
	0 & 0 & 0 & 0 & 0 & -(n-2)\mu-\lambda
	\end{bmatrix},\\
A_2 &= \begin{bmatrix}
	\lambda & 0 & 0 & 0 & 0 & 0\\[0.3em]
	0 & \lambda & 0 & 0 & 0 & 0\\[0.3em]
	0 & 0 & \lambda & 0 & 0 & 0\\[0.3em]
	0 & 0 & 0 & \lambda & 0 & 0\\[0.3em]
	0 & 0 & 0 & 0 & \lambda & 0\\[0.3em]
	0 & 0 & 0 & 0 & 0 & \lambda\\[0.3em]
	\end{bmatrix},\\
A_0 + A_1 + A_2 &= \begin{bmatrix}
	-n & 0 & n-2 & 2 & 0  & 0\\[0.3em]
	n-1 & -n & 0 & 0 & 1 & 0\\[0.3em]
	0 & n & -n & 0 & 0 & 0\\[0.3em]
	0 & n-2 & 0 & -(n-1) & 0 & 1\\[0.3em]
	0 & 0 & 0 & n-1 & -(n-1) & 0\\[0.3em]
	n-2 & 0 & 0 & 0 & 0 & -(n-2)
	\end{bmatrix} \mu
\end{align*}
Then,
\begin{align*}
x &= \begin{bmatrix}
	x_1 & x_2 & x_3 & x_4 & x_5 & x_6
	\end{bmatrix}
	= \begin{bmatrix}
	\frac{(n-2)(n^2-2n+2)}{3n-2} & \frac{(n+1)(n-2)^2}{3n-2} & \frac{(n-2)^2(n^2-2n+2)}{n(3n-2)} & n-2 & \frac{(n+1)(n-2)^2}{(n-1)(3n-2)} & 1
	\end{bmatrix}\\
\Rightarrow xA_0e^T &= (n-2)\mu(x_1 + x_4 + x_6), xA_2e^T = \lambda^*(x_1 + x_2 + x_3 + x_4 + x_5 + x_6)\\
\Rightarrow \lambda^* &= \frac{n(n-1)(n-2)(n^3-n^2+n-2)}{3n^5-12n^4+22n^3-29n^2+26n-8}\\
\Rightarrow \lambda_{\text{MDS}} - \lambda^* &= \frac{n}{3} - \frac{n(n-1)(n-2)(n^3-n^2+n-2)}{3n^5-12n^4+22n^3-29n^2+26n-8} = \frac{n}{3}\frac{2(2n^3-4n^2+n+2)}{3n^5-12n^4+22n^2-29n^2+26n-8} = \frac{n}{3}\frac{1}{O(n^2)}.
\end{align*}
}
\end{IEEEproof}

\end{document}